\documentclass{aa}

  \usepackage[latin1]{inputenc}
  \usepackage{graphicx}
  \usepackage{amsmath,amssymb}

  \newcommand{\msol}{$\rm M_\odot$}
  \newcommand{\jpb}[1]{\textup{#1}}
  \newcommand{\ed}[1]{\textup{#1}}
  \newcommand{\sue}[1]{\textup{#1}}
  \newcommand{\chris}[1]{\textup{#1}}

\begin{document}

  %==========================================================================
  %                           Titre et abstract
  %==========================================================================

% TITRE
%------
 \title{ISM Properties in Low-Metallicity Environments}
\subtitle{II. The Dust Spectral Energy Distribution of NGC~1569}
\titlerunning{The Dust Spectral Energy Distribution of NGC~1569}

% AUTEURS
%--------
\author{Fr\'ed\'eric~Galliano\inst{1} \and 
        Suzanne~C.~Madden\inst{1} \and
        Anthony~P.~Jones\inst{2} \and 
        Christine~D.~Wilson\inst{3} \and
        Jean-Philippe~Bernard\inst{2,4} \and
        Francine~Le~Peintre\inst{2,5}}
\authorrunning{F. Galliano et al.}
\institute{Service d'Astrophysique, CEA/Saclay, L'Orme des Merisiers,
           91191 Gif sur Yvette, France \and
           Institut d'Astrophysique Spatiale (IAS),
           Universit\'e de Paris XI, 91405 Orsay, France \and
           Department of Physics and Astronomy, McMaster University,
           Hamilton, ON L8S 4M1, Canada \and
           Centre d'\'Etude Spatial des Rayonnements (CESR), 31028 
           Toulouse, France \and
	   deceased October 2001}

% ABSTRACT
%---------
\abstract{
We present new 450 and 850~$\mu m$ SCUBA data of the dwarf galaxy
NGC~1569. 
We construct the mid-infrared to millimeter SED of NGC~1569, using 
ISOCAM, ISOPHOT, IRAS, KAO, SCUBA and MAMBO data, and model the SED in order 
to explore the nature of the dust in low metallicity environments. 
The detailed modeling is performed in a self-consistent way, synthesizing the 
global ISRF of the galaxy using an evolutionary synthesis model with further 
constraints provided by the observed MIR ionic lines and a photoionisation 
model. 
Our results show that the dust properties are different in this low metallicity
galaxy compared to other more metal rich galaxies.  
The results indicate a paucity of PAHs probably due to the destructive effects 
of the ISRF penetrating a clumpy environment and a size-segregation of grains
where the emission is dominated by small grains of size $\sim 3\;\rm nm$, 
consistent with the idea of shocks having a dramatic effect on the dust 
properties in NGC~1569.  
A significant millimetre excess is present in the dust SED which can be 
explained by the presence of ubiquitous very cold dust ($\rm T = 5-7\; K$). 
This dust component accounts for 40 to 70 $\%$ of the total dust mass in
the galaxy ($1.6 - 3.4 \times 10^5$~\msol )
\sue{and could be distributed in small clumps 
(size $\simeq$ a few pc) throughout the galaxy.
We find a gas-to-dust mass ratio of 740 - 1600, larger than that of the Galaxy
and a dust-to-metals ratio of $1/4$ to $1/7$.
We generate an extinction curve for NGC~1569, consistent with the modeled dust
size distribution.
This extinction curve has relatively steep FUV rise and smaller 2175~\AA\ 
bump, resembling the observed extinction curve of some regions in the Large
Magellanic Cloud.}
% MOTS-CLES
%----------
\keywords{ISM: dust, extinction --
          Galaxies: dwarf --
          Galaxies: starburst --
          Infrared: galaxies --
          Submillimeter}
}

\date{Received / Accepted}

\offprints{fredg@discovery.saclay.cea.fr}

\maketitle

  %==========================================================================
  %                          Introduction
  %==========================================================================

\section{Introduction}

The absorption of stellar radiation by dust and its subsequent
reemission in the infrared (IR) to the submillimeter (submm) is a fundamental 
process controlling the heating and cooling of the interstellar medium (ISM;
e.g. Tielens \& Hollenbach 1985). 
The dust IR spectral energy distribution (SED) of a galaxy is its footprint 
reflecting fundamental physical parameters such as initial mass function (IMF),
age, stellar population and metallicity. 
Thus, knowledge of the physical characteristics of dust in galaxies opens the 
door to understanding the star formation history and the evolution of
galaxies. 
However, an accurate interpretation of the SEDs of galaxies requires knowledge
of the detailed macroscopic and/or microscopic dust properties in galaxies, 
such as composition, emissivity, dust size distribution, and spatial 
distribution. 
How these properties are effected by variations in the metallicity of the ISM 
is the focus of our present study.

Dwarf galaxies in our local universe are ideal laboratories for
studying the interplay between the ISM and star formation in low-metallicity 
environments (Hunter \& Gallagher 1989). 
They are at relatively early epochs of their chemical evolution, possibly 
resembling distant protogalaxies in their early stages of star formation. 
The subject of dust formation in primordial environments
and the processes controlling the subsequent evolution of dust compels us to 
study the dust properties in nearby dwarf galaxies, with the eventual goal of 
constructing templates that would apply to primordial galaxies. 
While only a relatively small number of metal-poor galaxies has been observed 
at mid-infrared (MIR) wavelengths using the Infrared Space Observatory (ISO), 
it is already apparent that the characteristics of the MIR dust components 
differ remarkably from those of more metal-rich starburst galaxies (e.g. 
Lutz et al. 1998; Crowther et al. 1999; Thuan et al. 1999; Madden 2000;
Vacca et al. 2002; Madden et al. 2003). 
Is it correct to simply assume Galactic dust properties when modeling the dust 
in dwarf galaxies?

Discrepancies are arising as to the quantity of dust that is actually
present in even the lowest metallicity galaxies that are observed
today. 
Since optical observations are more readily accessible than MIR to far infrared
(FIR) wavelengths which require space and air borne platforms, the dust 
opacity has mostly been determined through the effects of obscuration, not 
from the dust emission properties.
These different measurements are not always consistent. 
Even at 1/50 solar metallicity, I~Zw~18 contains a non-negligible amount of 
reddening, determined from optical observations to be equivalent to 
$\rm A_v = 0.5$ (Cannon et al. 2002). In the dwarf galaxy SBS~0335-052 
(1/40 solar metallicity), $\rm A_v$ values as high as 20 to 30 have been 
\sue{suggested from MIR observations (Thuan et al. 1999; 
Plante \& Sauvage 2002), contrary to the low values, $\rm A_v < 1$, obtained 
by Dale et al. (2002) and Izotov et al. (1997).}
All of the details that go into modeling the dust properties in dwarf galaxies 
are critical for determining the quantity of dust that resides in these 
metal-poor galaxies. 
This must also be reconciled with current dust evolution models before we
can understand the early epoch of star formation.

We focus on the detailed modeling of the SEDs of dwarf galaxies,
constructing the SEDs with as much wavelength coverage as possible. 
We began with a sample of dwarf galaxies observed in the MIR (Madden et
al. 2000; Madden et al. 2003). 
IRAS and ISOPHOT observations are included, along with our recently-obtained 
ground-based submm (Galliano et al. 2002) and millimetre (mm) observations 
(Galliano et al. 2003).  
The basis of the dust modeling is the model of D\'esert et al. (1990; hereafter
called DBP90). 
To synthesise an input interstellar radiation field (ISRF), we use a stellar 
evolution model (P\'EGASE; Fioc \& Rocca-Volmerange 1997) and additional 
constraints obtained from a photoionisation model (CLOUDY; Ferland 1996). 
Here we present the complete modeling for NGC~1569, a nearby 
($D = 2.2 \pm 0.6$~Mpc; Israel 1988) dwarf galaxy with an average metallicity 
of 1/4 solar (Gonz\'alez Delgado et al. 1997), currently in the aftermath of a
massive burst of star formation (Israel 1988; Waller 1991). 
A dust SED model has recently been presented by Lisenfeld et al. (2002).
Recent Chandra observations find large inhomogeneities in the metal abundances 
with ranges from 0.1~Z$_\odot$ to 1~Z$_\odot$ (Martin et al. 2002), evidence 
that the ISM has been affected by numerous supernovae explosions since the 
last burst. 
NGC~1569 contains two bright super-star-clusters (SSCs) (Arp \& Sandage 1985; 
Aloisi et al. 2001) which have blown out a large fraction of the gas in a
dramatic display.
  
The paper is organised as follows.
Section~\ref{sec:obs} presents an overview of our new
observations and the data we use from the literature.
Section~\ref{sec:model} describes the adopted model and the method we use to
investigate the dust properties.
Section~\ref{sec:results} presents the modeled SED of NGC~1569 and the 
consequences of the results on the dust properties.
We end with a summary and the conclusions in Sect.~\ref{sec:concl}.

  %==========================================================================
  %                 Observations et reduction de donnees
  %==========================================================================

\section{The Observations}
\label{sec:obs}

% On fait reference
%------------------
An observed SED was constructed, as completely as possible, incorporating
data from the literature for various telescopes: IRAS, KAO, ISO
(ISOPHOT) and IRAM (MAMBO), as well as our ISOCAM data (Madden et
al. 2003) and our new 450 and 850~$\mu m$ JCMT (SCUBA) observations presented
here. 
Since we are modeling the global SED in this paper, the differences in beam 
sizes are not of great concern here.

      %------------------------------------------------------------------
      %                       Donnees SCUBA
      %------------------------------------------------------------------

  \subsection{SCUBA observations}
  \label{sec:scuba}

% Presentation de SCUBA
%----------------------
We obtained 450 and 850 $\mu m$ data of NGC~1569, with SCUBA (Holland et al. 
1999), a bolometer array on the James Clerk Maxwell Telescope (JCMT), during 
two observing runs in February 2000 and December 2000.
Observations were carried out in the jiggle-mapping mode using a 64-point 
jiggle pattern with a chop throw of $150''$.

% Reduction des donnees
%----------------------
We reduced these data using the SURF V1.5-1 software (Jenness \& Lightfoot 
1998), other STARLINK utilities such as KAPPA, CONVERT, FIGARO and our own 
IDL routines. 
We scrutinised the data between each step of the processing to check for 
possible artifacts. 
We proceeded in the following way:
\\
  % Correction de l'extinction 
%--------------------------- 
1) Atmospheric extinction: 
The relatively low atmospheric transmission is the main difficulty in 
obtaining good quality ground-based observations at these wavelengths. 
To quantify the atmospheric extinction and correct our data for this effect, 
we performed skydips at 450~$\mu m$ and 850~$\mu m$ every 1-2 hours. 
In addition, we used the measured $\tau_{\rm CSO}$ at 225 GHz, from the Caltech
Submillimeter Observatory radiometer, which ranged from 0.04 to 0.1 during our 
observations. 
We evaluate the atmospheric transmission at 850~$\mu m$, $\tau_{850}$, using 
the skydip value adapted to the elevation of the bolometer. 
However, the 450~$\mu m$ skydip has been determined to be unreliable due to 
the fluctuations of the atmosphere. 
We, therefore, used relationships between $\tau_{\rm CSO}$, $\tau_{850}$ and 
$\tau_{450}$ to compute $\tau_{450}$: 
$\tau_{450} = 6.52\times (\tau_{850}-0.049)$ (Archibald et al. 2000).
\\
  % Despiking et babdbol
  %---------------------
2) Flatfields, despiking and bad bolometers: 
The non-uniformity of the response of each bolometer is corrected with 
standard calibration flatfield measurements since it remains constant with 
time and does not need to be re-measured every night (Jenness \& Lightfoot 
1998). 
The spikes were removed from individual maps using a $4\sigma$ threshold 
and noise measurements were performed at the beginning and at the end of the
shift every night. 
Bad bolometers were removed with a threshold of $3\sigma$. 
This last step can create holes in individual maps. 
However, since the array rotates on the sky, the final map contains no holes.
\\
  % Fluctuations du ciel
  %---------------------
3) Sky fluctuations: 
The fluctuations in the emissivity of the atmosphere were removed using the 
fact that the outer ring of bolometers was observing the sky.
The median of the outer ring, excluding the noisy bolometers, was subtracted 
from each map. 
Using cumulative growth-curves, we verified that no source flux was removed. 
Each map was rebinned with a pixel field of view of $6''$/pixel at 850~$\mu m$
and $3''$/pixel at 450~$\mu m$ which correspond to the jiggle-map sampling.
\\
  % Calibration
  %------------
4) Flux conversion: 
Each night, we performed three measurements of the calibration sources, Uranus,
Mars or CRL~618, using the same observing parameters. 
An average Flux Conversion Factor (FCF) was computed from these measurements, 
for each night and the individual maps were calibrated by the corresponding 
FCF. 
We evaluated the average peak-to-aperture ratios, $R_{\rm date}(\lambda)$, on 
the calibrators, to correct for the loss of flux in secondary lobes. 
The full-width-half maximum (FWHM) beam size is $8.5''$ at 450~$\mu m$ and 
$15.2''$ at 850~$\mu m$.
\\
  % Rebin
  %------
5) Map coadding and filtering: 
The individual maps were coadded, weighted by $T_{\rm int}/\sigma^2$, where 
$T_{\rm int}$ is the integration time and $\sigma$ the standard deviation of 
the pixels in the image. 
We applied the MR/1 multiresolution filtering method (Starck et al. 1998) 
using a $3\sigma$ threshold.
\\
  % Erreurs
  %--------
6) Evaluation of uncertainties: 
To estimate the cumulative uncertainty $\Delta F_\nu (\lambda)$ on the flux 
$F_\nu (\lambda)$ integrated in a circular aperture $\Theta$ \chris{(we took 
$\Theta = 120''$ for the entire galaxy)}, we quantified the individual error 
contributions. 
We first construct RMS maps for 850~$\mu m$ and 450~$\mu m$ observations 
such that each pixel $(i,j;\lambda)$ of these maps is the standard deviation 
$\sigma_{\rm RMS}(i,j;\lambda)$ of the corresponding pixels of the individual 
maps. 
The contribution of these fluctuations to the total error,
$\Delta F_\nu^{\rm RMS}(\lambda)$, is computed by integrating the flux in the 
RMS map in the aperture $\Theta$: $\Delta F_\nu^{\rm RMS}(\lambda) =
 \sqrt{\sum_{(i,j)\in\Theta}
 (\sigma_{\rm RMS}(i,j;\lambda)R_{\rm date}(\lambda)N_{\rm beam})^2}$
where $N_{\rm beam}$ is the number of pixels per beam. 
The error due to the sky emission subtraction, $\sigma_{\rm sky}$, is the 
standard deviation of the distribution of the points used to compute the 
median: 
$\Delta F_\nu^{\rm sky} (\lambda) =
\sqrt{N_\Theta(\sigma_{\rm sky}R_{\rm date}(\lambda)N_{\rm beam})^2}$ where 
$N_\Theta$ is the number of pixels inside the aperture $\Theta$. 
The error on the absolute calibration, considering the distribution of the 
FCF, $\{C_{\rm cal}(\lambda)\}$, and the distribution of the 
peak-to-aperture ratios, $\{R_{\rm cal}(\lambda)\}$: 
$\Delta F_\nu^{\rm conv}(\lambda) =
 \sigma (C_{\rm cal}(\lambda))/
 \langle C_{\rm cal}(\lambda)\rangle\times F_\nu (\lambda)$,
$\Delta F_\nu^{\rm ratio}(\lambda) =
 \sigma (R_{\rm cal}(\lambda))/
 \langle R_{\rm cal}(\lambda)\rangle\times F_\nu (\lambda)$.
The total uncertainty on the net flux is the sum of these different
contributions (Table~\ref{tab:errSCUBA}):
\begin{eqnarray}
  \Delta F_\nu (\lambda) & = & \left[
                               (\Delta F_\nu^{\rm RMS} (\lambda))^2
                               + (\Delta F_\nu^{\rm sky} (\lambda))^2
                               \right. \nonumber \\
                         & + & \left.
                               (\Delta F_\nu^{\rm conv}(\lambda))^2
                               + (\Delta F_\nu^{\rm ratio} (\lambda))^2
                               \right]^{1/2} .
\end{eqnarray}
\begin{table}[h]
  \begin{center}
    \begin{tabular}{l*{5}{c}}
    \hline
    \hline
      \multicolumn{1}{c}{}    & \bf RMS  & \bf Sky
      & \bf Calibrator & \bf Ratio & \bf Total       \\
    \hline
      % 850 microns
      $\pmb{850 \mu m}$        & $8\,\%$  & $5\,\%$
      & $5\,\%$        & $4\,\%$   & $\pmb{12\,\%}$  \\
      % 450 microns
      $\pmb{450 \mu m}$        & $38\,\%$ & $4\,\%$
      & $20\,\%$       & $14\,\%$  & $\pmb{45\,\%}$  \\
    \hline
    \end{tabular}
    \caption{The various contributions to the uncertainties on the net flux of
             the SCUBA maps integrated
             in a $120''$ diameter aperture centered at
             $\alpha(2000) = 4^{\rm h}30^{\rm m}49^{\rm s}$,
             $\delta(2000) = 64^\circ 50'55''$.}
    \label{tab:errSCUBA}
  \end{center}
\end{table}
\\
  % Contributions autres 
  %--------------------- 
7) Radio and molecular contributions: 
The radio continuum emission contaminating the 450~$\mu m$ and 850~$\mu m$ 
broadbands was evaluated from Israel \& de Bruyn (1988).  
Taking their value of the spectral index, $\alpha = -0.36$, we extrapolate 
the radio continuum to be 49~mJy ($\sim 14\,\%$) at 850~$\mu m$ and 39~mJy 
($\sim 3\,\%$) at 450~$\mu m$. 
We estimate the CO(3-2) contribution in the observed 850~$\mu m$ band from 
Meier et al. (2001) to be $\sim 5\,\%$. 
The observed 450~$\mu m$ and 850~$\mu m$ flux values we use for our dust 
modeling were adjusted for non-dust contamination.

  % Lisenfeld et al.
  %-----------------
Our 850~$\mu m$ value \sue{($345\pm 40$~mJy) is consistent with Lisenfeld et 
al. (2002) ($410\pm 45$~mJy)} using different SCUBA data.
Our 450~$\mu m$ flux \sue{is $1320\pm 450$~mJy while Lisenfeld et al. 
(2002) report a higher total flux of $1820\pm 700$~mJy but still consistent.}
By studying the behavior of both the 450 and 850~$\mu m$ skydips as a 
function of time of day and sky conditions, we gave particular attention to 
calibration questions. 
The most accurate calibration uses the 850~$\mu m$ skydips to adjust the 
450~$\mu m$ maps.  
Using the 450~$\mu m$ skydip to calibrate the 450~$\mu m$ data is known to
give higher fluxes (Archibald et al. 2000). 
This is probably the greatest source of difference in the final calibration 
between the present paper and Lisenfeld et al. (2002).

      %------------------------------------------------------------------
      %                     Donnees de la litterature
      %------------------------------------------------------------------

  \subsection{Infrared to millimetre data}

\begin{figure*}[htbp]
  \begin{center}
    \begin{tabular}{cc}
      \includegraphics[width=9cm]{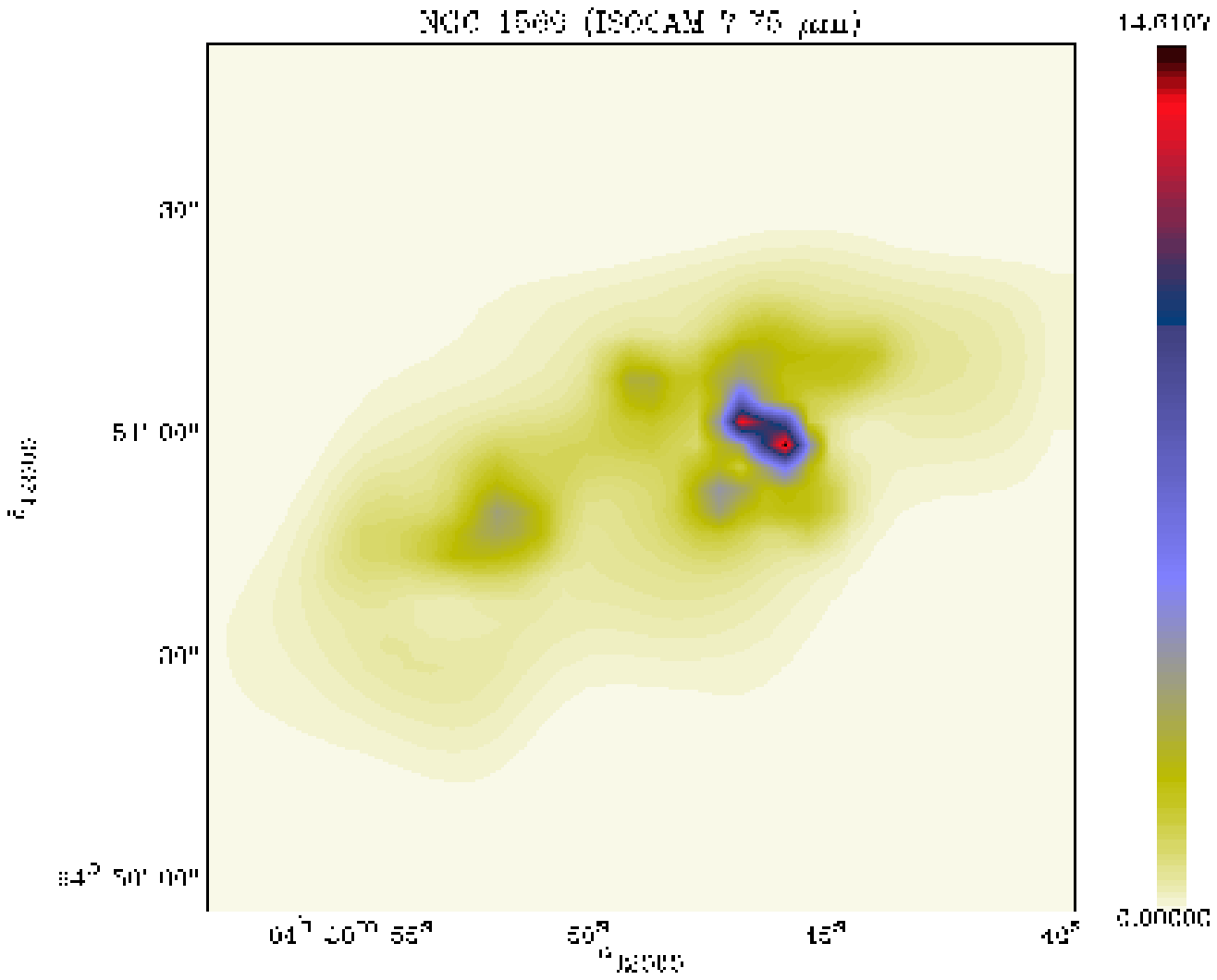} & \hspace{-2cm}
      \includegraphics[width=9cm]{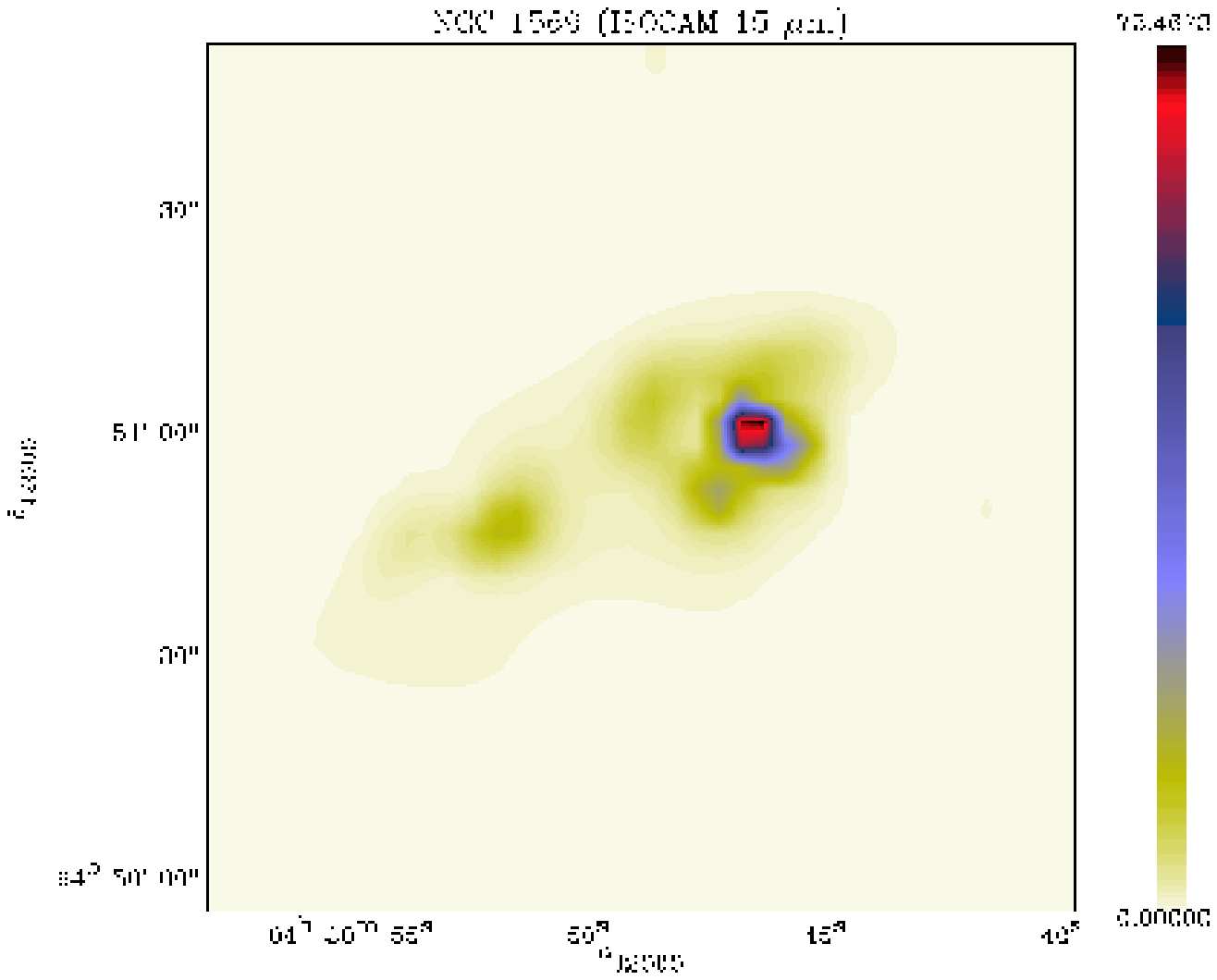} \\
      \includegraphics[width=9cm]{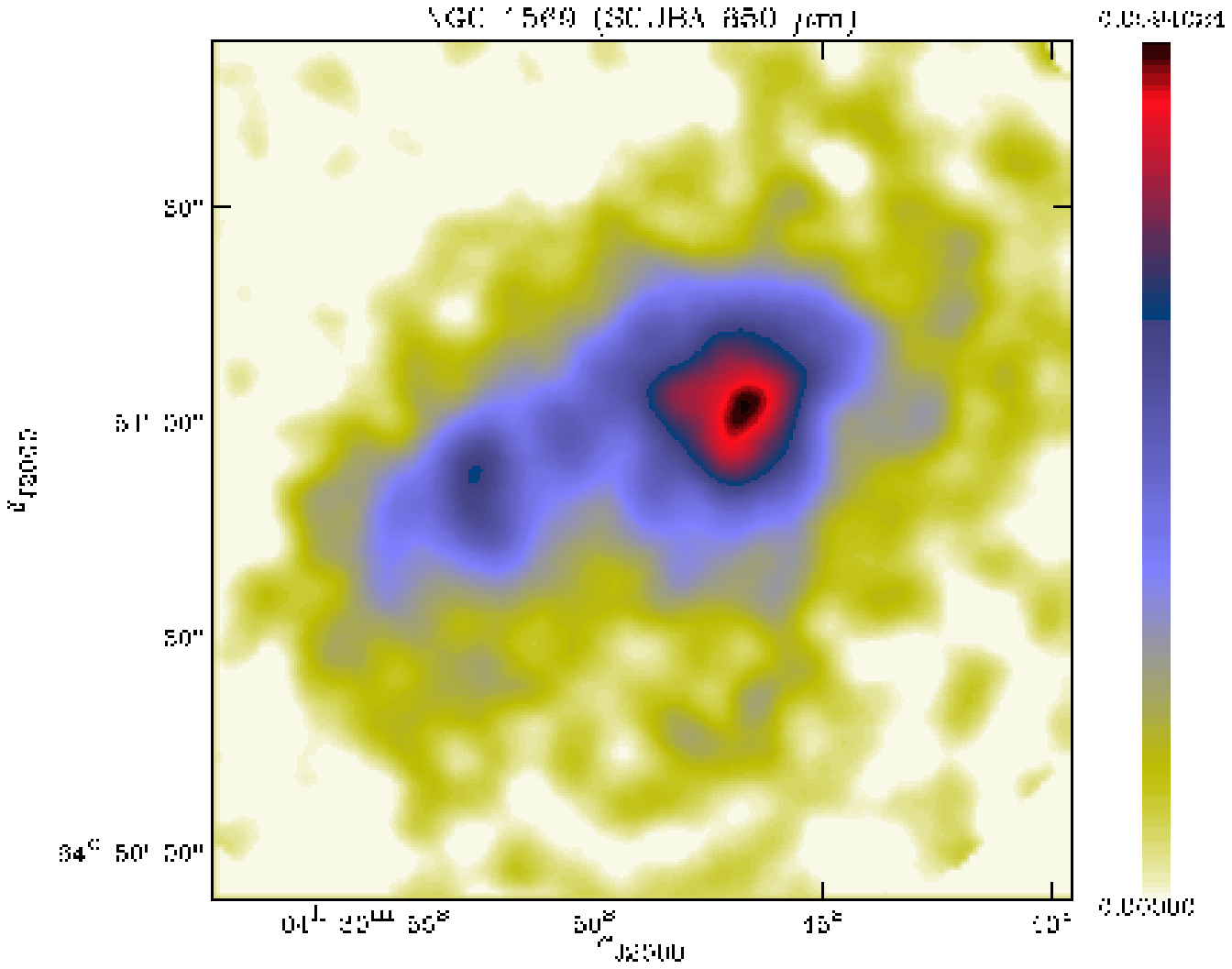}  & \hspace{-2cm}
      \includegraphics[width=9cm]{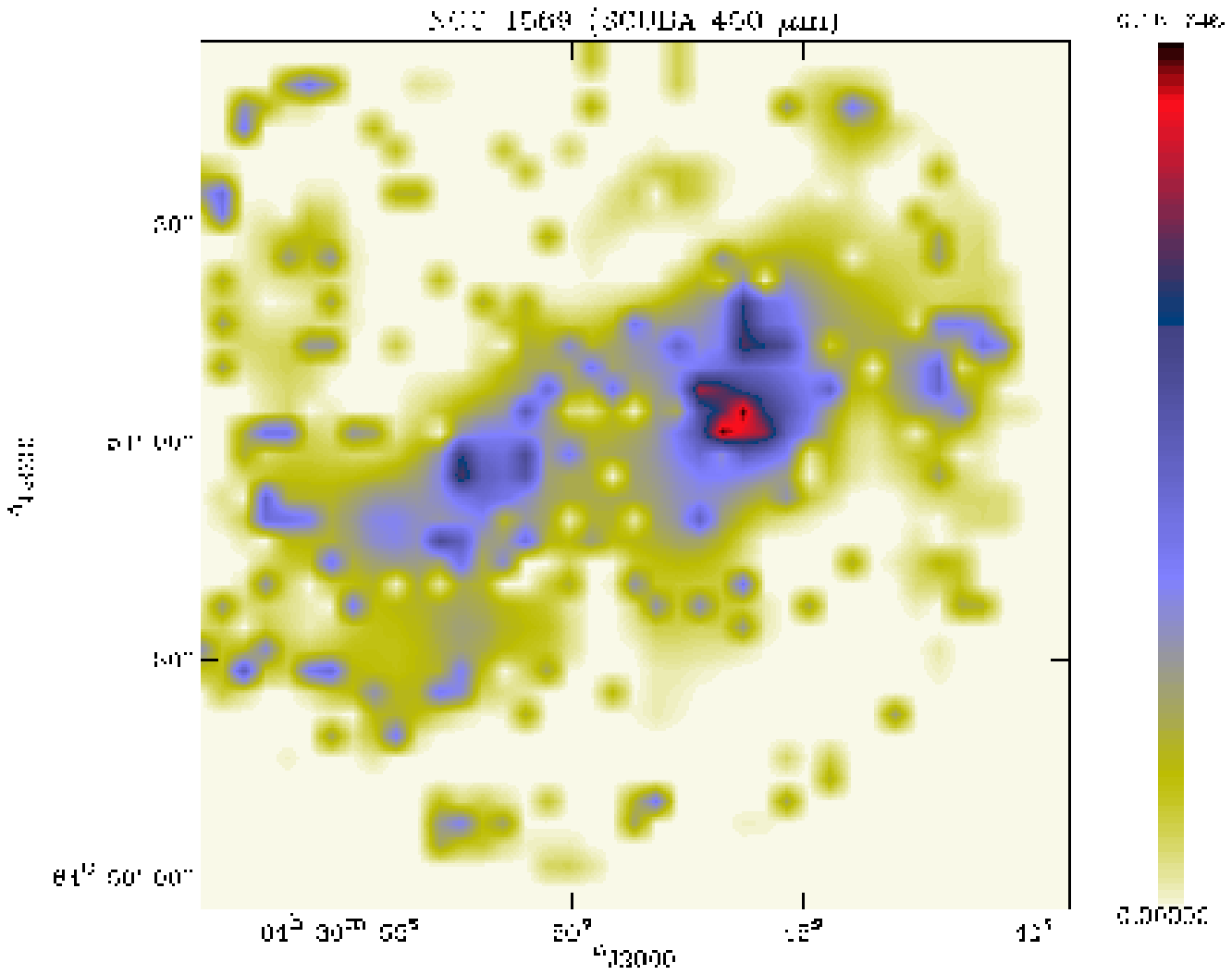}  \\
    \end{tabular}
    \caption{ISOCAM and SCUBA images of NGC~1569.
             Top-left: ISOCAM-LW6 ($\lambda = 7.75\; \mu m$, beam=$7.6''$),
             this band traces PAH features.
             Top-right: ISOCAM-LW3 ($\lambda = 15\; \mu m$, beam=$9.9''$),
             this band traces the hot dust continuum.
             Bottom-left: SCUBA ($\lambda = 850\; \mu m$, beam=$15.2''$).
             Bottom-right: SCUBA ($\lambda = 450\; \mu m$, beam=$8.5''$).
             These two submm bands trace the cold dust continuum.
             The field of view is the same for the 4 images and the color table
             has the same dynamic range. 
             The ISOCAM images have been deconvolved.}
    \label{fig:images}
    \begin{tabular}{cc}
      \includegraphics[width=9cm]{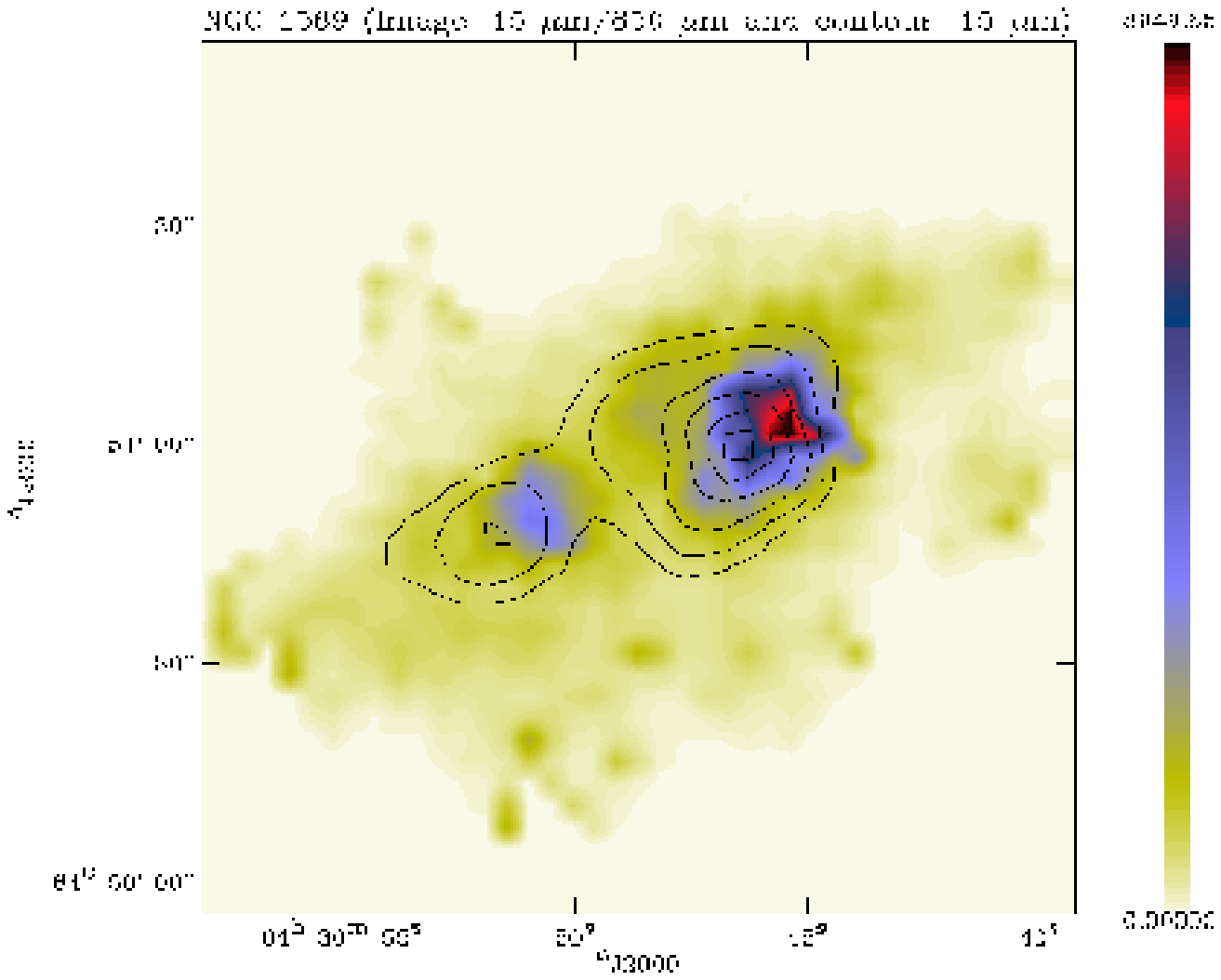} & \hspace{-2cm}
      \includegraphics[width=9cm]{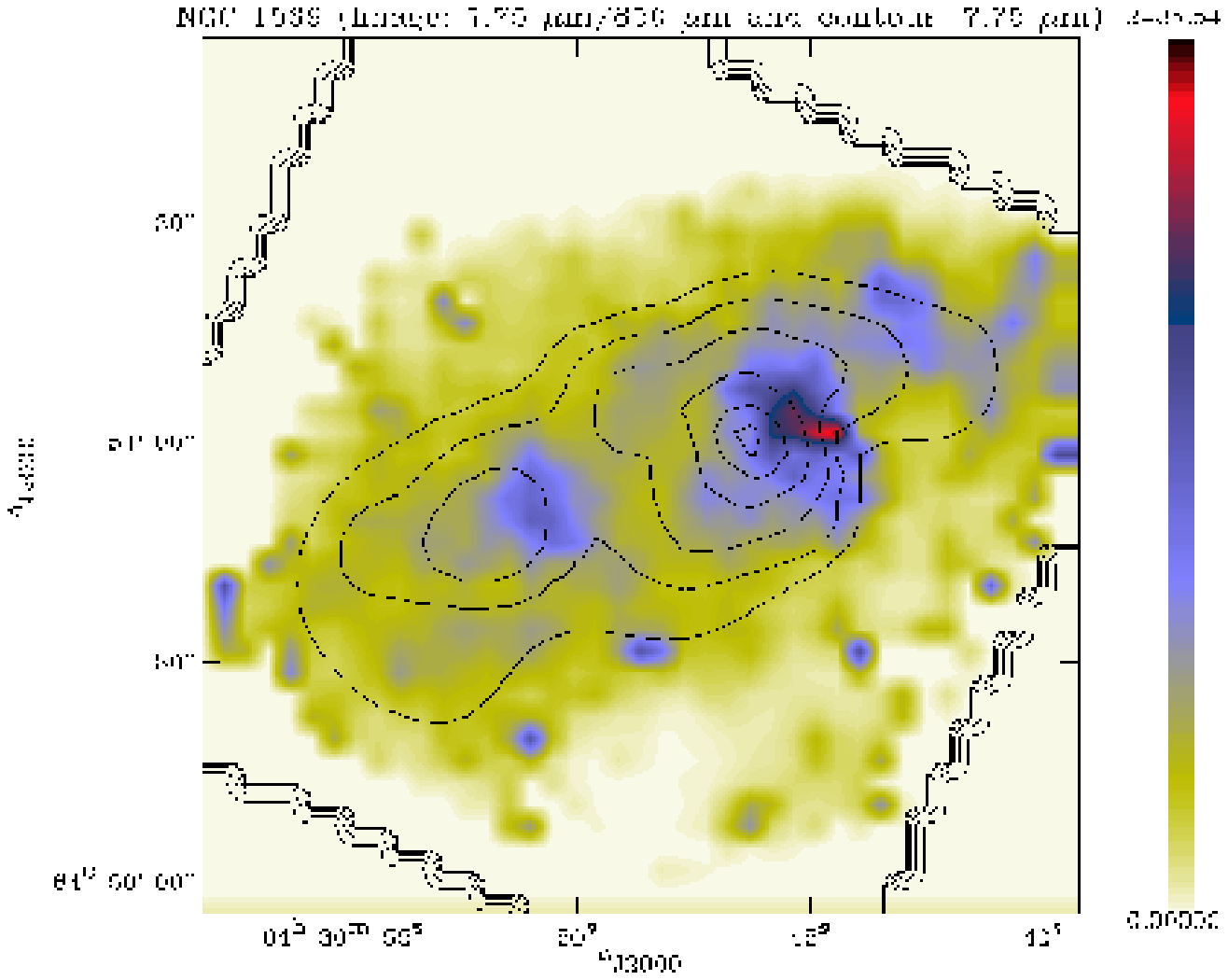} \\
    \end{tabular}
    \caption{Ratio of hot dust/cold dust.
             Left: ratio of the ISOCAM-LW3 image
             (contours, $\lambda = 15\; \mu m$) to the SCUBA image
             ($\lambda = 850\; \mu m$).
             Right: ratio of the ISOCAM-LW6 image
             (contours, $\lambda = 7.75\; \mu m$) to the SCUBA image
             ($\lambda = 850\; \mu m$).
             The ISOCAM images have been degraded to the resolution of the
             SCUBA (850~$\mu m$) image ($15''$).}
    \label{fig:rapCAM_850}
  \end{center}
\end{figure*}

% Le CVF et les images
%---------------------
The ISOCAM circular variable filter (CVF) spectrum provides the
constraints in the 5 to 16~$\mu m$ wavelength range. 
The details and the data treatment of the CVF observations are presented in 
Madden et al. (2003). 
We have found that the slope of the MIR CVF is a critical factor in 
constraining the dust model. 
Thus, using only the IRAS 12~$\mu m$ band or several ISOCAM broad band
observations do not provide sufficient constraints on the model. 
We characterised the MIR dust continuum by choosing 10 wavelength regions of 
the CVF which do not contain aromatic bands or ionic lines
(Table~\ref{tab:IRflux}).  
This was to balance the full observed SED which contains 8 additional points 
from other instruments. 
The ISOCAM images (Fig.~\ref{fig:images}) are $2 \times 2$ raster maps and 
have been processed in the same way as the CVF spectrum (Madden et al. 2003). 
We deconvolved the individual images using a multiresolution Lucy algorithm 
from the MR/1 package (Starck et al. 1998), using a $3\sigma$ detection 
threshold.
Some ratio maps are shown in Fig.~\ref{fig:rapCAM_850}.

% Au sujet des flux IRAS
%-----------------------
IRAS fluxes for NGC~1569 are given by Thronson \& Telesco (1986), Hunter et 
al. (1989a) and Melisse \& Israel (1994). 
The values for the four IRAS broadbands given by Thronson \& Telesco (1986) 
and by Hunter et al. (1989a) are similar, and equal to the ISOCAM CVF when
integrated over the equivalent IRAS 12~$\mu m$ band. 
The values reported by Melisse \& Israel (1995) differ significantly and are 
flagged by the authors to be uncertain. 
For our modeling purposes, we use the values of Hunter et al. (1989a) since 
error bars are also provided (Table~\ref{tab:IRflux}). 
These fluxes have been color-corrected.

% Les donnees ISOPHOT
%--------------------
ISOPHOT is an imaging polarimeter operating between 2.5 and 240 $\mu m$, which
was onboard the satellite ISO (Lemke et al. 1996). 
Data were taken by the two cameras of PHOT-C, C\_100 and C\_200, at 60, 100 
and 180 $\mu m$.
However after analysing the data, we realised that the background position 
was on a bright cirrus spot, rendering the data difficult to interpret. 
Instead, we use only the value from the ISOPHOT 170~$\mu m$ serendipity 
survey (Stickel et al. 2000) which gives a total 170~$\mu m$ flux of 
$(28 \pm 9.8)$~Jy for NGC~1569 (Martin Haas, private communication).

% Flux IRAM et KAO
%-----------------
Hunter et al. (1989b) measured a 155~$\mu m$ flux of 36~Jy on the Kuiper 
Airborne Observatory (KAO). 
The 1.2~mm IRAM flux is from Lisenfeld et al. (2002). 
They have subtracted the contributions of the radio continuum and the 
CO(2-1) line emission. 
All of the fluxes used are listed in table~\ref{tab:IRflux}.
%CHANGE
\begin{table}[h]
  \begin{center}
    \begin{tabular}{*{4}{l}}
     \hline
     \hline
       \bf Instrument & $\pmb{\lambda\; (\mu m)}$ 
       & $\pmb{\Delta\lambda\; (\mu m)}$ & \bf Net flux (mJy)      \\
     \hline
       ISOCAM         & 8.8          & 0.05          & $(370\pm 220)^a$   \\
                      & 10.1         & 0.96          & $(480\pm 230)^a$   \\
                      & 10.8         & 0.87          & $(610\pm 210)^a$   \\
                      & 11.8         & 0.73          & $(860\pm 220)^a$   \\
                      & 12.1         & 0.69          & $(980\pm 220)^a$   \\
                      & 13.2         & 0.54          & $(1070\pm 220)^a$  \\
                      & 13.9         & 0.45          & $(1270\pm 230)^a$  \\
                      & 14.6         & 0.35          & $(1420\pm 240)^a$  \\
                      & 15.0         & 0.30          & $(1470\pm 250)^a$  \\
                      & 16.0         & 0.16          & $(1990\pm 290)^a$  \\
     \hline
       IRAS           & 24.3         & 6.3           & $(8550\pm 1100)^b$ \\
                      & 62.8         & 19            & $(48900\pm 7000)^b$\\
                      & 103          & 22            & $(54800\pm 8000)^b$\\
     \hline
       KAO            & 155          & 30            & $(36000\pm 11000)^c$\\
     \hline
       ISOPHOT        & 174          & 90            & $(28000\pm 9800)^d$\\
     \hline
       SCUBA          & 443          & 20            & $(1280\pm 450)^e$  \\
                      & 863          & 70            & $(280\pm 60)^e$    \\
     \hline
       IRAM           & 1200         & 200           & $(190\pm 60)^f$    \\
     \hline
    \end{tabular}
    \caption{Net fluxes for the \sue{dust emission by the} entire galaxy used 
             for the modeling.
             The values take account of all the corrections mentioned in the
             text.
             $^a$~Madden et al. (2003), $^b$~Hunter et al. (1989a),
             $^c$~Hunter et al. (1989b), $^d$~Stickel et al. (2000)
             \& full catalog in prep., $^e$~this paper, 
             $^f$~Lisenfeld et al. (2002).}
    \label{tab:IRflux}
  \end{center}
\end{table}

    % UV / OPTIQUE
    %-------------
    \subsection{UV and optical data}

To constrain the input stellar radiation field (Sect.~\ref{sec:isrf}) we used
optical and UV data measured for the total galaxy (Israel 1988; De
Vaucouleurs et al. 1991; Table~\ref{tab:UVflux}).
\begin{table}[h]
  \begin{center}
    \begin{tabular}{*{2}{l}}
     \hline
     \hline
       \bf Wavelength  & \bf Flux density        \\
       \bf (\AA)       & \bf ($\pmb 10^{-14}$ erg s$^{-1}$
                         cm$^{-1}$ \AA$^{-1}$)   \\
     \hline
       1500            & $(152\pm 48)^a$         \\
       1800            & $(208\pm 25)^a$         \\
       2200            & $(146\pm 25)^a$         \\
       2500            & $(100\pm 18)^a$         \\
     \hline
       3650            & $(108\pm 10)^b$         \\
       4400            & $(101\pm 9)^b$          \\
       5500            & $(70\pm 6)^b$           \\
     \hline
    \end{tabular}
    \caption{Fluxes for the entire galaxy used to model the global
             ISRF.
             The values are corrected for Galactic extinction.
             $^a$ Israel (1988), $^b$ De Vaucouleurs et al. (1991).}
    \label{tab:UVflux}
  \end{center}
\end{table}

% Page contenant les images
%--------------------------

  %==========================================================================
  %                         Description du modele
  %==========================================================================

\section{Self-consistent modeling of the global SED}
\label{sec:model}

% La methode
%-----------
We have modeled the IR to millimetre dust emission from NGC~1569 in a
self-consistent way using a wide variety of observational constraints.
In this section we describe the method used. 
The results are presented in Sect.~\ref{sec:results}.
The general algorithm used to obtain the best solution is outlined in 
Sec.~\ref{sec:iterative}.

    %----------------------------------------------------------------------
    %                      Les parametres observationnels
    %----------------------------------------------------------------------

  \subsection{Physical parameters of the galaxy}
  \label{sec:obspar}

\begin{figure}[h]
  \begin{center}
  \includegraphics[width=\hsize]{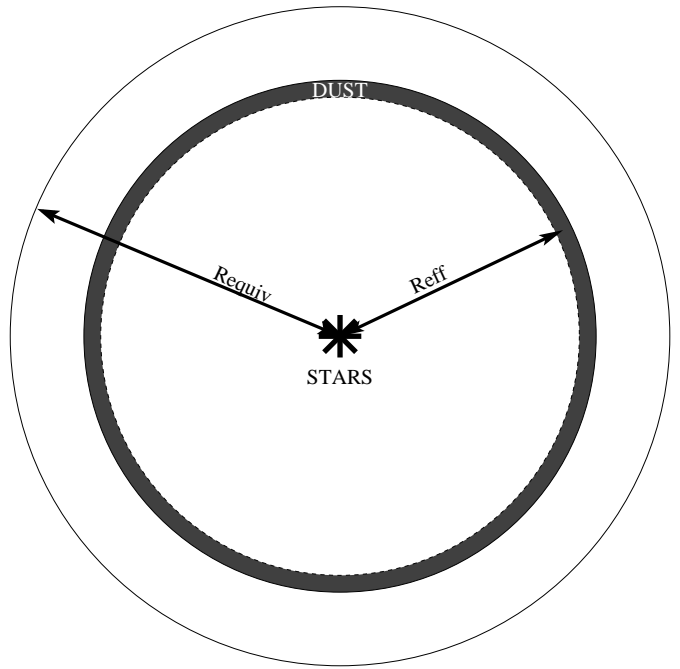}
  \caption{Schematic diagram of the model geometry.}
  \label{fig:geometry}
  \end{center}
\end{figure}

%CHANGE
%This model assumes spherical geometry and that all the dust is located at 
%an effective distance ($\rm R_{eff}$) from the central heating sources, the 
%stars. 
%It does not take into account the effects of radiative transfer.}
\ed{In the adopted model, we assume that the dust is located in a thin shell 
at an effective distance ($\rm R_{eff}$) from the central heating sources, 
the stars (Fig.~\ref{fig:geometry}). 
The extinction is treated in terms of a slab model which neglects scattering.}
%CHANGE

In order to have a consistent set of physical parameters (average
radius, density, column density, etc.), we estimate the radius $\rm R_{equiv}$
of the equivalent spherical galaxy \ed{from our 850~$\mu m$ image} and deduce
the value of the densities from this radius and from the hydrogen
masses given in the literature.
To determine $\rm R_{equiv}$, we fit an ellipse to the galaxy at 850~$\mu m$
using the geometrical mean of the axes as the 
average radius, which gives an equivalent radius, $\rm R_{equiv}$, of 
$0.65 \pm 0.18\; \rm kpc$ (the uncertainty, $\rm \Delta R_{equiv}$, is due 
to the uncertainty in the distance), consistent with the optical size: 
$1.85 \times 0.95$~kpc (Israel 1988) which gives a 
$\rm R_{equiv} = 0.66$~kpc.
The corresponding effective hydrogen density, $\rm n_{eff}$, and column
density, $\rm N_{eff}$, are $6.7$~cm$^{-3}$ and $1.8\times 10^{22}$~cm$^{-2}$ 
from the M(HI) and M(H$_2$) given by Israel (1997)
($\rm M(HI) = 1.4\times 10^8$~\msol, $\rm M(H_2) = 0.5\times 10^8$~\msol).

The mean distance from the dust to the stars is taken to be 
$\rm R_{eff} = 3/4 \times R_{equiv} = 0.50\; kpc$, corresponding to a 
homogeneous spatial distribution of the dust. 
We ran the model for this value of $\rm R_{eff}$  and also for two extreme 
cases: 
$\rm R_{eff}^{max} = R_{equiv} + \Delta R_{equiv} = 0.84\; kpc$ and 
$\rm R_{eff}^{min} = 1/2 \times R_{equiv} - \Delta R_{equiv} =
0.24\; kpc$. 
These extreme values permit us to investigate a wide variety of spatial
distributions and also to take into account uncertainties on the distance 
determination.

    %----------------------------------------------------------------------
    %                      Le modele de poussiere
    %----------------------------------------------------------------------

  \subsection{Dust modeling}
  \label{sec:dbp90}

% Description
%------------
To compute the dust emission spectrum we use DBP90 which includes three 
dust components and provides a coherent interpretation of both the 
interstellar extinction and the infrared emission. 
This model has been used to explain the dust properties in the Galaxy. 
The three components of the DBP90 model are the following:
\begin{enumerate}
  \item Polycyclic Aromatic Hydrocarbons (PAHs); 2-dimensional molecules
        responsible for the MIR emission features and the FUV non-linear
        rise in the extinction curve.
        In the standard DBP90 model applied to the Galaxy, PAHs have a size 
        range of $0.4\;\rm nm \lesssim a \lesssim 1.2\; nm$.
  \item Very Small Grains (VSGs); 3-dimensional carbonaceous grains
        responsible for the MIR continuum emission and the
        2175 \AA\ extinction bump.
        In the standard model, their radii are
        $1.2\;\rm nm \lesssim a \lesssim 15\; nm$.
  \item Big Grains (BGs); 3-dimensional silicates, coated or
        mixed with carbonaceous material, responsible for the FIR emission and
        for the NIR and visible rise of the extinction curve.
        In the standard model, their radii are
        $15\;\rm nm \lesssim a \lesssim 110\; nm$.
\end{enumerate}
Due to their small sizes, and thus low heat capacity, PAHs and VSGs are 
transiently heated by single photon absorption and are not in thermal 
equilibrium with the radiation field. 
BGs usually reach thermal equilibrium for the grain sizes determined for the 
Galaxy. 
However the dust model does include thermal fluctuations for the BGs if their 
sizes warrant it. 
The model also computes the absorption cross section of the grains 
corresponding to the physical parameters that fit the emission. 
Finally, the extinction curve is generated using the absorption cross sections
generated by the model.

% Utilisation
%------------
We use the DBP90 model to compute the MIR to millimetre SED. 
The data which constrain the SED are summarised in Table~\ref{tab:IRflux}. 
To obtain the best fit to these data, we have varied the physical 
parameters of the model, using a Levenberg-Marquardt method 
(Numerical Recipes, Press et al. 1996) to minimize the $\chi^2$. 
The parameters which can vary are:
\begin{enumerate}
\item the mass abundance of each species, $Y=m/m_{\rm H}$ ($m$ is the dust mass
      in the beam and $m_{\rm H}$ is the hydrogen mass
      in the beam);
\item the index of the power-law of the size distribution of each species,
      $\alpha$, where $n(a) \propto a^{-\alpha}$ ($a$ is the grain radius and
      $n(a)$ the number density of grains between $a$ and $a+da$);
\item the minimum and maximum sizes of each species, $a_{min}$ and $a_{max}$;
\item the absorption coefficient of the grains, $Q_{\rm abs}$, defined as
      $\sigma_{\rm abs} = Q_{\rm abs}\pi a^2$ ($\sigma_{\rm abs}$ is the
      absorption cross-section) for a spherical grain of radius $a$.
\end{enumerate}
Moreover, the widths of the individual broadband observations are taken into 
account, except for the ISOCAM CVF spectrum since the spectral resolution 
is sufficient to neglect this effect ($\rm R \geq 35$). 
For the data from the bolometers (SCUBA and MAMBO), we integrated the modeled 
SED into the observational broadbands. 
For the data from photo-multipliers (KAO and ISOPHOT),  we made the 
corresponding color-correction when integrating into the broadbands.
For the IRAS data, this correction was already made. 
This correction has the largest effect for the 170~$\mu m$ ISOPHOT broadband 
(Fig.~\ref{fig:dustembb}).

    %----------------------------------------------------------------------
    %                    Modelisation de l'ISRF
    %----------------------------------------------------------------------

  \subsection{Modeling of the interstellar radiation field}
  \label{sec:isrf}

The DBP90 model requires an input ISRF which
heats the dust. 
To be consistent, we have modeled the global ISRF for NGC~1569 using the 
UV and optical data (Table~\ref{tab:UVflux}) coupled with the evolutionary 
synthesis model P\'EGASE 2.0 (Fioc \& Rocca-Volmerange 1997), taking into 
account the constraints from photonisation processes using CLOUDY.

    % PEGASE
    %-------
    \subsubsection{Stellar evolutionary synthesis}

% Description
%------------
P\'EGASE computes the stellar spectral energy distribution taking into account
the metallicity evolution. 
%CHANGE
%We do not incorporate the dust extinction effects and the nebular emission 
%with P\'EGASE. 
%The dust extinction is computed by DBP90 and the nebular emission is taken 
%into account using the photoionisation model, CLOUDY 
%(Sect.~\ref{sec:cloudy}).
\ed{We do not incorporate the dust extinction effects and the nebular emission 
into P\'EGASE. 
The $Q_{\rm abs}$ values are computed by DBP90, the extinction is calculated 
with a simple screen model and the nebular emission is taken 
into account using the photoionisation model, CLOUDY 
(Sect.~\ref{sec:cloudy}).}

% Le fit
%-------
We fit the UV-to-optical data (Table~\ref{tab:UVflux}) using a combination of 
two single instantaneous bursts with an initial metallicity of 
$\rm Z=Z_\odot/4.27$ (Gonz\'alez Delgado et al. 1997). 
The reddening is calculated using the dust cross-sections computed by the 
dust model. 
However, the optical depth deduced from this extinction curve corresponds to 
the case where all the dust is located in front of the stars. 
To obtain the effective optical depth, $\tau_\nu^{\rm eff}$, the shape of the
extinction curve is scaled in order to satisfy the energy conservation:
%CHANGE
%\begin{equation}
%  \rm L_\star = L_{UV-opt} + L_{IR-mm}
%\end{equation}
%where
%\begin{equation}
%  \left\{
%  \begin{array}{l}
%    \displaystyle \rm L_\star = 4\pi R_{eff}^2\,
%                  \int_{-\infty}^{+\infty} I_\nu^{stell}\, d\nu \\
%    \displaystyle \rm L_{UV-opt} = 4\pi R_{eff}^2\,
%                  \int_{-\infty}^{+\infty} I_\nu^{stell}\,
%                                    \exp(-\tau_\nu^{eff})\, d\nu \\
%    \displaystyle \rm L_{IR-mm} = 4\pi R_{eff}^2\,
%                  \int_{-\infty}^{+\infty} I_\nu^{dust}\, d\nu
%  \end{array}
%  \right.
%\end{equation}
%and $\rm I_\nu^{stell}$ and $\rm I_\nu^{dust}$ are the monochromatic
%intensities emitted by the stars and the dust in a sphere of radius
%$\rm R_{eff}$.
\ed{
\begin{equation}
  \rm F_\star = F_{UV-opt} + F_{IR-mm}
\end{equation}
where $\rm F_\star$ is the flux density as it would be observed if there were 
no dust in the galaxy, only stars, and $\rm F_{UV-opt}$ and $\rm F_{IR-mm}$ 
are the flux densities from reddened stars and dust respectively, as they
would be observed:
\begin{equation}
  \left\{
  \begin{array}{l}
    \displaystyle \rm F^{UV-opt}_\nu =  F^\star_\nu \exp(-\tau_\nu^{eff}) \\
    \displaystyle \rm F_{UV-opt} = 
       \int_{-\infty}^{+\infty}  F^\star_\nu \exp(-\tau_\nu^{eff})\, d\nu \\
    \displaystyle \rm F_{IR-mm} = 
       \int_{-\infty}^{+\infty}  F^\star_\nu \left[1 - 
                                 \exp(-\tau_\nu^{eff})\right]\, d\nu \, .
  \end{array}
  \right.
\end{equation}
}

We minimise the $\chi^2$ to find the best age combination. 
The fit gives several age combinations with similar $\chi^2$. 
To remove this degeneracy, we use a further constraint, comparing the MIR 
ionic line ratios from Madden et al. (2003) and the theoretical results 
computed with the photoionisation model CLOUDY (version 90.04, Ferland 1996).

    % CLOUDY
    %-------
    \subsubsection{Photoionisation}
    \label{sec:cloudy}

% Description
%------------
CLOUDY predicts the spectra of astrophysical plasma in different environments.
It uses a recent atomic database and takes into account the geometry and the 
gas properties.
We use CLOUDY through the MICE\footnote{MICE is supported by the SWS and the 
ISO Spectrometer Data Center at MPE through DLR (DARA) under grants 50 QI 
86108 and 50 QI 94023.} IDL interface.
%CHANGE
\ed{A thick shell geometry is adopted while that for the dust emission is a 
thin shell.}

% La grille de parametres
%------------------------
We have computed a grid of solutions from CLOUDY containing a range of values 
for the following parameters:
\begin{enumerate}
  \item the shape of the input ISRFs are those which give the best $\chi^2$
        values as the output from P\'EGASE;
  \item the total luminosity of the galaxy was computed by integrating the 
        total UV-optical and IR-mm SED: 
        $\rm L_{tot} = 1.9\times 10^9\; L_\odot$;
  \item the inner radius varies from $\rm R_{\rm in}=5\; pc$ to
        $\rm R_{\rm in}=90\; pc$, which is the radius of the hole in HI 
        centered on the SSC A (Israel \& Van Driel 1990);
  \item the average hydrogen number density is $\rm n(H)_{eff}$
        ($6.7$~cm$^{-3}$; Sect.~\ref{sec:obspar});
  \item the outer radius was taken to be $\rm R_{equiv}$
        (Sect.~\ref{sec:obspar}).
\end{enumerate}
We set the elemental abundances to $X=X_{\rm H}=X_\odot$, 
$Y=X_{\rm He}=Y_\odot$ and $Z=1-X-Y=Z_\odot/4.27$ according to the value of 
the metallicity given by Gonz\'alez Delgado et al. (1997).
Fitting the theoretical ionic line ratios to the MIR data 
($[\rm NeIII]/[\rm NeII]$ = $9.6 \pm 1.4$; 
$[\rm SIV]/[\rm NeIII]$ = $0.49 \pm 0.08$; Madden et al. 2003) clearly
removes the degeneracy of the ISRF.

    %----------------------------------------------------------------------
    %                        Processus iteratif
    %----------------------------------------------------------------------

  \subsection{Iterative process}
  \label{sec:iterative}

\begin{figure}[h]
  \begin{center}
  \includegraphics[width=\hsize]{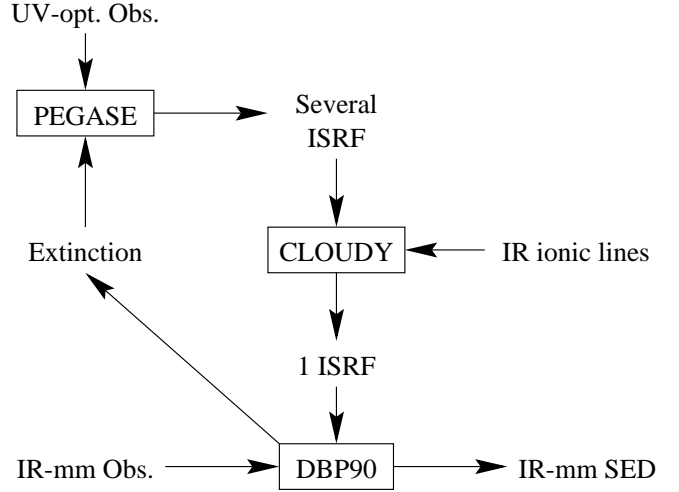}
  \caption{Algorithm used to compute the SED in a self-consistent way.
           ``UV-opt. Obs.'' are the data contained in Table~\ref{tab:UVflux},
           ``IR-mm Obs.'' are the data contained in Table~\ref{tab:IRflux},
           ``IR ionic lines'' are from Sect.~\ref{sec:cloudy},
           ``IR-mm SED'' is the modeled dust SED and
           ``Extinction'' is the extinction curve computed by DBP90 after
           converging to a solution.
           We iterate on this process to obtain the necessary consistency 
           between emission and extinction.}
  \label{fig:method}
  \end{center}
\end{figure}

% Schema general
%---------------
The modeling of the dust emissivity generates an extinction curve which is 
used to deredden the UV-to-optical data in order to compute the ISRF. 
The ISRF is used by the dust emissivity model to heat the dust. 
Therefore, to be consistent, we compute this sequence through an iterative 
process until we reach a stable, self-consistent solution. 
This solution is obtained when the extinction curve used to deredden the
data equals that deduced from the dust emission. 
The general scheme of this iterative process is shown in 
Fig.~\ref{fig:method}.
% Algorithme (Schema)
%--------------------

  %==========================================================================
  %                          Resultats et discussion
  %==========================================================================

\section{Results and discussion}
\label{sec:results}

We have applied the general model described in Sect.~\ref{sec:model} to our
observed IR-to-millimetre SED for NGC~1569 (Table~\ref{tab:IRflux}).

    %----------------------------------------------------------------------
    %                Les resultats sur l'emission des poussieres
    %----------------------------------------------------------------------

  \subsection{The dust emissivity}

In this subsection we describe the details of the solution of the dust 
modeling and discuss the constraints and the consequences of the model 
results.

    % Exces millimetrique
    %--------------------
    \subsubsection{Millimetre excess}
    \label{sec:millimex}

\begin{figure*}[ht]
  \begin{center}
  \includegraphics[width=\textwidth]{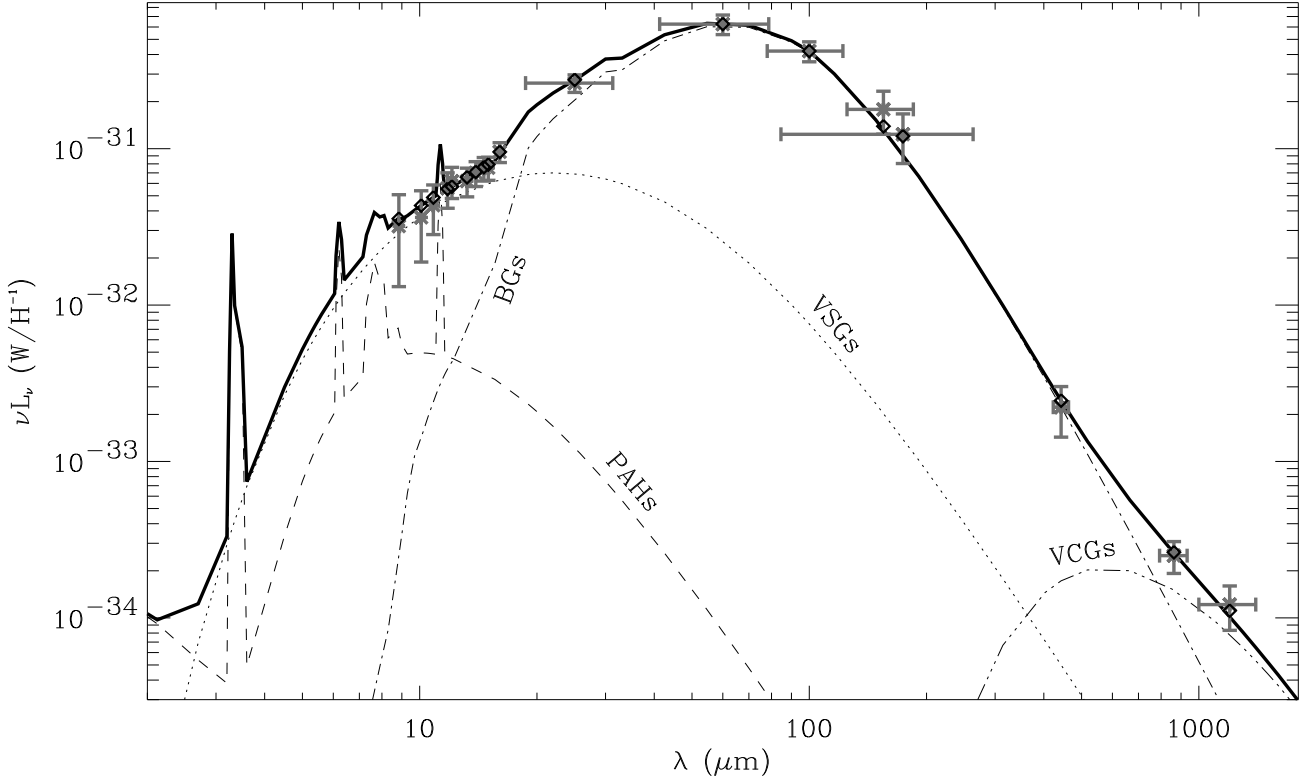}
  \caption{NGC~1569 observations and modeled SED.
           The data (Table~\ref{tab:IRflux}) are indicated by crosses: 
           vertical bars are the errors on the flux values and the horizontal 
           bars indicate the widths of the broadbands.
           The lines are the dust model and its different components 
           (see Sect.~\ref{sec:millimex}).
           Diamonds are the model integrated over the observational broadbands 
           and color-corrected.
           Thus, departures from the model lie where the diamonds deviate 
           from the crosses.
           The power is expressed in $\rm W\, H^{-1}$ which is $\rm \nu L_\nu$
           divided by the number of H atoms ($\rm L_\nu$ is the monochromatic
           luminosity).}
  \label{fig:dustembb}
  \end{center}
\end{figure*}

% Le probleme
%------------
In areas where we have no strong constraints, some parameters of the general
model described in Sect.~\ref{sec:dbp90} are kept fixed, while others are
left as free parameters to obtain the best fit.
\begin{enumerate}
  \item PAHs: this component is too weak in NGC~1569
        to be constrained with our set of data.
        An upper limit on the mass abundance of the PAHs,
        $Y_{\rm PAH}$, is fixed and we used the Galactic values for $\alpha$, 
        $a_{\rm min}$ and $a_{\rm max}$.
  \item VSGs: we varied the mass abundance, $Y_{\rm VSG}$, and two
        parameters governing the size distribution, $\alpha$ and $a_{\rm max}$.
        However, we cannot constrain $a_{\rm min}$ since
        the $\chi^2$ process tends to decrease it to very small sizes which 
        have no physical meaning, so it was kept constant at the Galactic 
        value.
  \item BGs: we varied the mass abundance, $Y_{\rm BG}$, and two
        parameters governing the size distribution, $\alpha$ and $a_{\rm min}$.
        Since the results are not sensitive to $a_{\rm max}$, we adopt
        the Galactic value.
\end{enumerate}
We are able to fit the SED quite well except in the millimetre wavelength 
range, where there is an excess that can not be explained by the three 
standard DBP90 components (see Fig.~\ref{fig:dustembb}). 
%We are unable to obtain better than $\chi^2 = 10$ for 11 degrees of freedom 
%with the three component model.
% Figure de la SED
%------------------
\begin{figure*}[ht]
  \begin{center}
  \includegraphics[width=\textwidth]{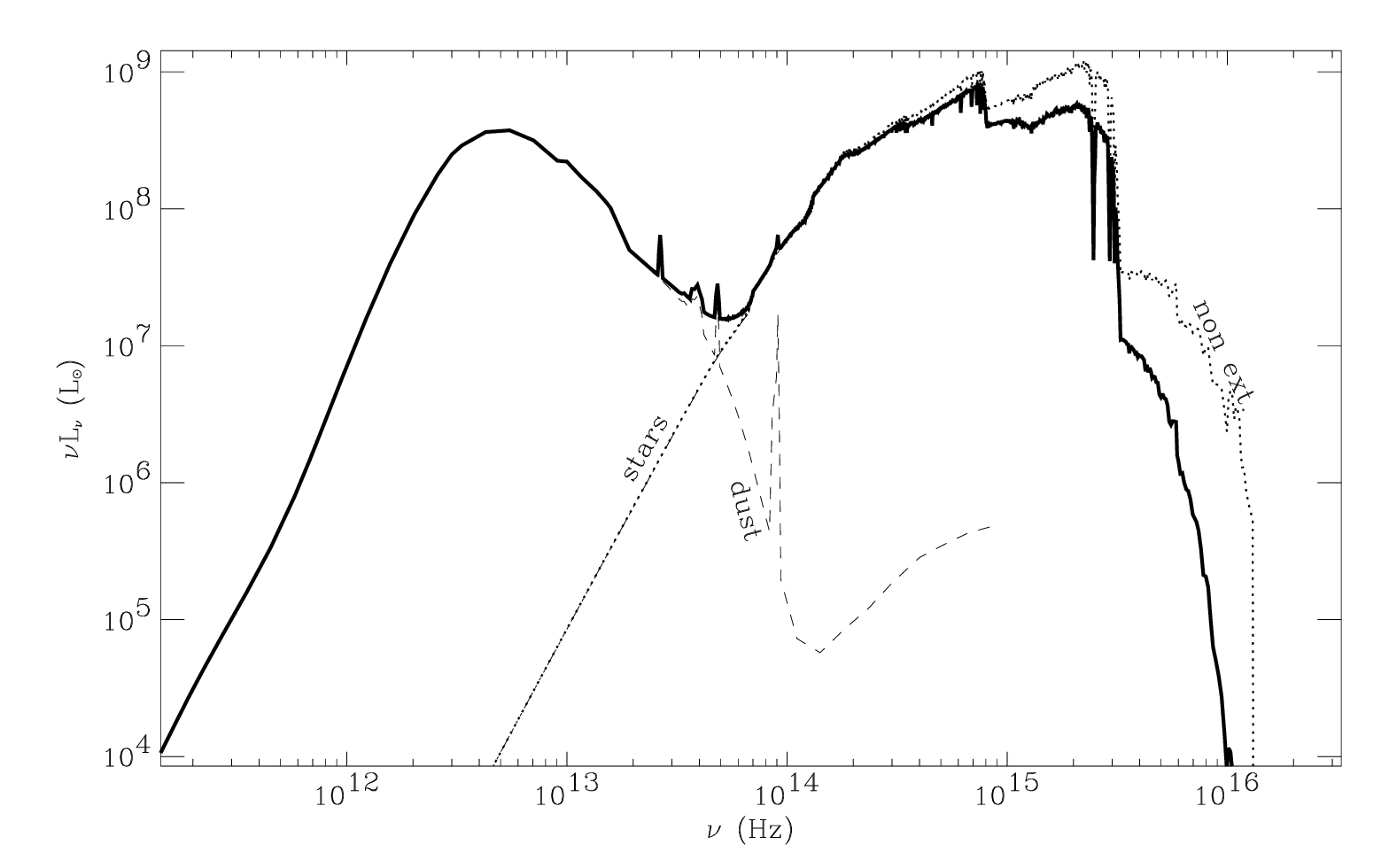}
  \caption{Global synthesized SED for NGC~1569.
           The solid line is the total energy of the galaxy: the sum of the
           dust (dashed line) plus the emerging stellar radiation 
           (dashed-dotted line).
           For comparison, the dotted line shows the non-extincted stellar 
           radiation.}
  \label{fig:sed_tot}
  \end{center}
\end{figure*}

% La composante froide
%---------------------
\chris{Although very cold dust is not the only likely explanation of this
excess,
we have added a fourth component modeled with a modified black-body to 
describe it.
We call this component Very Cold Grains (VCGs). 
We discuss the other possible origins and the consequences of this 
excess in Sect.~\ref{sec:explsub}}.
The monochromatic flux radiated by the VCG component in a sphere of radius 
$\rm R_{eff}$ is given by:
\begin{equation}
  \rm
  F_\nu^{\rm VCG} =
  \frac{\displaystyle 3 Y_{\rm VCG} M(H)}
       {\displaystyle 4 \rho R_{eff}^2}
  \left(\displaystyle \frac{Q_0}{a_0}\right)
  \left(\displaystyle \frac{\lambda_0}{\lambda}\right)^\beta
  \frac{2hc/\lambda^3}{\exp(hc/\lambda kT)-1}
\end{equation}
where $\rm M(H)$ is the total mass of hydrogen in the galaxy, $\lambda_0$,
%CHANGE
\ed{$\rm Q_0$ and $\rho$ are the reference wavelength, the absorption 
coefficient at $\lambda_0$ and the mass density of the grains.} 
This VCG radiated flux depends on three variables: the temperature, $T$, the 
emissivity index, $\beta$, and the mass abundance of the grains, 
$Y_{\rm VCG}$ (we adopted typical values: 
$\rm Q_0/a_0 = 40\; cm^{-1}$, $\lambda_0 = 250 \;\mu m$, 
$\rho = \rho_{\rm silicate} = 3.3 \;\rm g\, cm^{-3}$ if $\beta = 2$ and 
$\rho = \rho_{\rm graphite} = 2.2 \;\rm g\, cm^{-3}$ if $\beta = 1$).
As we have only three data points to constrain the submillimetre-millimetre
wavelengths regime of the SED (450~$\mu m$, 850~$\mu m$ and 1.2~mm), we 
consider a way to fix the value of one parameter to avoid the degeneracy.  
Since the best solution requires a relatively flat slope, we fix $\beta = 1$ 
for the $\chi^2$ solution. 
Setting $\beta = 2$ gives us an unphysical solution with a maximum temperature 
of 3 K.
Our emissivity index of 1 is lower than that found by Dunne \& Eales (2001) for
a sample of IR-bright galaxies ($\beta \simeq 2$). 
However, the assumption of a lower $\beta$ is reasonable in light of the fact 
that Agladze et al. (1996) measure the absorption coefficient of cosmic dust 
analog grains and find that over the temperature range of 1.2-10~K, $\beta$ 
decreases with temperature. 
Thus, the mass abundance, $Y_{\rm VCG}$, and temperature, T, remain the only 
free parameters governing the VCG properties.

    % Contraintes sur les prop.
    %--------------------------
    \subsubsection{What constraints can we put on the dust properties~?}
    \label{sec:dusterr}

% Resultat du fit
%----------------
Our best fit of the observed SED is shown in Fig.~\ref{fig:dustembb}.
The values of the parameters corresponding to the best $\chi^2$ fit for the 4
component model are summarized in Table~\ref{tab:param}.
\begin{table*}
\begin{center}
  \begin{tabular}{l|*{3}{ll|}ll}
    \hline
    \hline
      \multicolumn{1}{c|}{}              &
      \multicolumn{2}{c|}{\bfseries PAH} &
      \multicolumn{2}{c|}{\bfseries VSG} &
      \multicolumn{2}{c|}{\bfseries BG}  &
      \multicolumn{1}{c}{} & \multicolumn{1}{l}{\bfseries VCG} \\
    \hline
      \multicolumn{1}{c|}{}             &
      \bf Milky       & \bf NGC         &
      \bf Milky       & \bf NGC         &
      \bf Milky       & \bf NGC         &
                      & \bf NGC         \\
      \multicolumn{1}{c|}{}             &
      \bf Way         & \bf 1569        &
      \bf Way         & \bf 1569        &
      \bf Way         & \bf 1569        &
                      & \bf 1569        \\
    \hline
      $\pmb Y$        &
      $4.3\times 10^{-4}$ & $\lesssim 1.0\times 10^{-6}$ &
      $4.7\times 10^{-4}$ & $1.8\times 10^{-5}$ &
      $6.4\times 10^{-3}$ & $4.4\times 10^{-4}$ &
      $\pmb Y$            & $(1.3 - 0.4)\times 10^{-3}$ \\
    \hline
      $\pmb a_-$      &
      4 \AA           & 4 \AA     &
      12 \AA          & 12 \AA    &
      150 \AA         & 22 \AA    &
      \bf T           & $5 - 7\; K$       \\
    \hline
      $\pmb a_+$      &
      12 \AA          & 12 \AA    &
      150 \AA         & 78 \AA    &
      1100 \AA        & 1100 \AA  &
      $\pmb \beta$    & 1.0             \\
    \hline
%    \cline{8-9}
      $\pmb \alpha$   &
      3               & 3               &
      2.6             & 4.0             &
      2.9             & 6.3             &
      \multicolumn{2}{c}{} \\
    \cline{1-7}
  \end{tabular}
  \caption{The values are those for the best fit.
           For comparison, we give the corresponding values for the Galaxy 
           from D\'esert et al. (1990).}
  \label{tab:param}
\end{center}
\end{table*}
This corresponds to $\chi^2=1.5$ with 9 degrees of freedom.
% General
%--------
In order to quantify the reliability of the grain properties deduced from our 
model, we attempt to estimate, as conservatively as possible, the errors on 
these parameters.

\begin{figure}[htb]
  \begin{center}
  \includegraphics[width=\hsize]{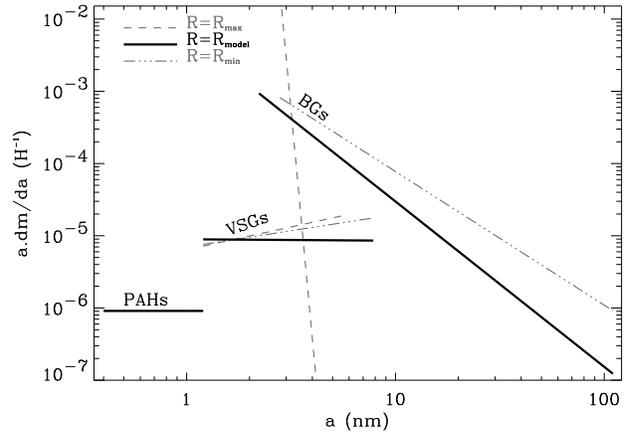}
  \caption{Grain size distribution for the two extreme radius cases
           ($\rm R_{eff}^{max}$ and $\rm R_{eff}^{min}$)
           and for the best estimated value, $\rm R_{eff}$ 
           (see section~\ref{sec:obspar}).
           It gives the conservative range within which the size distribution 
           can be constrained.}
  \label{fig:spread_dist}
  \end{center}
\end{figure}

We investigated the two possible sources of uncertainties: the sensitivity of 
the model itself to certain parameters, and to the value of the average 
radius ($\rm R_{eff}$) which influences the radiation field intensity.
% Observations
%-------------
First, we estimate the errors on the goodness of the fit, by quantifying the 
departure of the model from the observations.
We independently vary each parameter to fit the extreme cases allowed by the
error bars.
For example, the VSGs are constrained mainly by the observations from 8 to 
16~$\mu m$. 
Thus, we made 4 extreme fits of the VSGs: one through the 
upper edges of the error bars, one through the lower edges, one through the
upper edge of the error bar at 8~$\mu m$ and the lower edge at 16~$\mu m$,
and one through the lower edge at 8~$\mu m$ and the upper edge at 16~$\mu m$.
This gives us an idea of the sensitivity of the model to the physical 
properties.
% Hypothese sur R
%----------------
Second, the model was solved for the two extreme values of the radius ($\rm
R_{eff}^{max}$ and $\rm R_{eff}^{min}$ defined in 
Sect.~\ref{sec:obspar}). 
These two extreme cases give us an idea of the influence of the geometry 
(and also of the ISRF) on the dust properties. 
Figure~\ref{fig:spread_dist} demonstrates the effect of the assumed radius on 
the grain size distributions. 

Table~\ref{tab:dusterr} contains the ranges of the physical parameters for 
the dust components given the sources of uncertainty from the fitting 
and the geometry.
The PAH and VSG dust parameters are not very sensitive to the assumed 
geometry but are relatively sensitive to the fitting.
%CHANGE
%, an effect related to the stochastic heating nature of these grains. 
For these cases, $a_{max}$ can change, at most, by a factor of 3. 
The best estimate $\rm R_{eff}$ shows a flat size distribution for the VSGs, 
and a small slope for the 2 extreme geometry cases 
(Fig.~\ref{fig:spread_dist}). 
%CHANGE
%The slope of the BG size distribution, however, is very sensitive to the
%assumed geometry if $\rm R_{eff}$ is decreased to $\rm R_{eff}^{min}$. 
%In this extreme case, the slope of the size distribution decreases 
%substantially, leaving no ``big grains'' and the size distribution of the BG 
%component overlaps with that of the VSGs.
\ed{The slope of the BG size distribution, however, is very sensitive to the
assumed geometry.
If $\rm R_{eff}$ is increased to $\rm R_{eff}^{max}$, the size distribution 
shows an approximately constant size of $\sim 3$~nm (shown as the near 
vertical dashed line in Fig.~\ref{fig:spread_dist}), with a deficit of larger 
grains.
When the dust is located farther from the stars, the size of the 
grains must be smaller to reach the same temperature, which is 
constrained by the observations.}
\begin{table}
  \begin{center}
    \begin{tabular}{llll}
      \hline
      \hline
        \multicolumn{2}{c}{} & \multicolumn{1}{l}{\bf Lower limit}
                             & \bf Upper limit  \\
      \hline
        \bf PAH  & $Y$       & $0$              & $1.0\times 10^{-6}$  \\
      \hline
        \bf VSG  & $\alpha$  & $2.6$            & $5.2$            \\
                 & $a_+$     & $35$ \AA         & $120$ \AA        \\
                 & $Y$       & $1.3\times 10^{-5}$  & $2.3\times 10^{-5}$  \\
      \hline
        \bf BG   & $\alpha$  & $5.8$            & $35$             \\
                 & $a_-$     & $21$ \AA         & $29$ \AA         \\
                 & $Y$       & $3.5\times 10^{-4}$  & $4.7\times 10^{-4}$  \\
      \hline
    \end{tabular}
    \caption{Range of reliability of the physical parameters of the grains
             (from errors in the fitting and variations in the assumed 
              geometry).}
    \label{tab:dusterr}
  \end{center}
\end{table}

% Temperature des VCGs
%---------------------
We treat the sensitivity of the parameters of the VCGs separately. 
Indeed, the temperature of the VCGs is a very important parameter, which has 
major consequences for the derived dust mass, and is only constrained by two 
data points.
%CHANGE******** 
\jpb{In this case, the temperature is strongly dependent on the emissivity 
index, however we excluded $\beta=2$ in Sect.~\ref{sec:millimex} to avoid 
unphysical solutions.}
The temperature does not depend on the radiation field since we added this 
fourth component as a modified blackbody. 
The best fits within 1~$\sigma$ variation give a range of 
temperatures: $\rm 5\; K \lesssim T \lesssim 7\; K$.
The abundances corresponding to this temperature range are 
$0.4\times 10^{-3} \lesssim Y_{\rm VCG} \lesssim 1.3\times 10^{-3}$.

    % Proprietes des poussieres
    %--------------------------
    \subsubsection{The dust properties - what are the consequences?}
    \label{sec:dustprop}

\begin{figure*}[ht]
  \begin{center}
  \includegraphics[width=\textwidth]{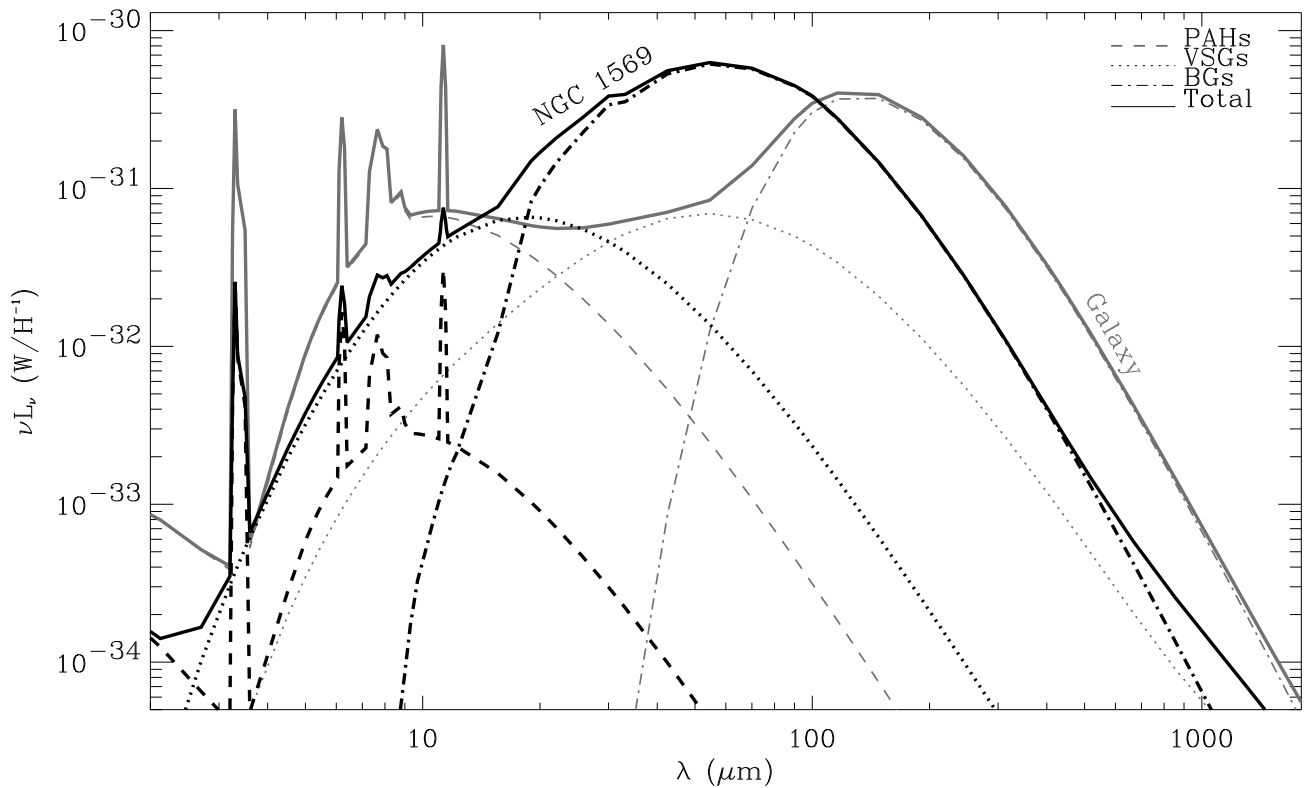}
  \caption{Comparison of the dust SED for NGC~1569 (black) with the Galaxy 
           (grey, \chris{from DBP90}).
           In addition to the total dust SED (solid lines), the
           individual components of the DBP90 dust model
           also shown for both the Galaxy and NGC~1569.
           The power is expressed in $\rm W\, H^{-1}$ which is $\rm \nu L_\nu$
           divided by the number of H atoms ($\rm L_\nu$ is the monochromatic
           luminosity).}
  \label{fig:compdustSED}
  \end{center}
\end{figure*}

% Description de l'emission
%--------------------------
The dust properties deduced from this study differ from those of normal, more 
metal rich galaxies like the Milky Way.
If we compare the shape of the dust SED of NGC~1569 to the SED of the Galaxy
(Fig.~\ref{fig:compdustSED}), we notice the lack of PAHs.  
Moreover, the VSG component in the Galaxy and in NGC~1569 emit roughly similar
energies, as do the BG components in both galaxies. 
However, this energy is emitted at shorter wavelengths in NGC~1569 than in 
the Galaxy since these grains are in a more energetic environment. 
On the long-wavelength side of the SED, the energy emitted from the Galaxy is 
$\sim 10$ times higher than that of NGC~1569.
Even though the excess in the submm does not contribute much to the global
energy budget of NGC~1569, most of the dust mass resides in this excess while
in the Galaxy most of the dust mass is in the BG component.

      % Distribution de taille
      %-----------------------
      \subsubsection*{The dust size distribution and extinction curve}

These differences noted above are reflected in the size distribution of the 
grains.
\begin{figure}[htb]
  \begin{center}
  \includegraphics[width=\hsize]{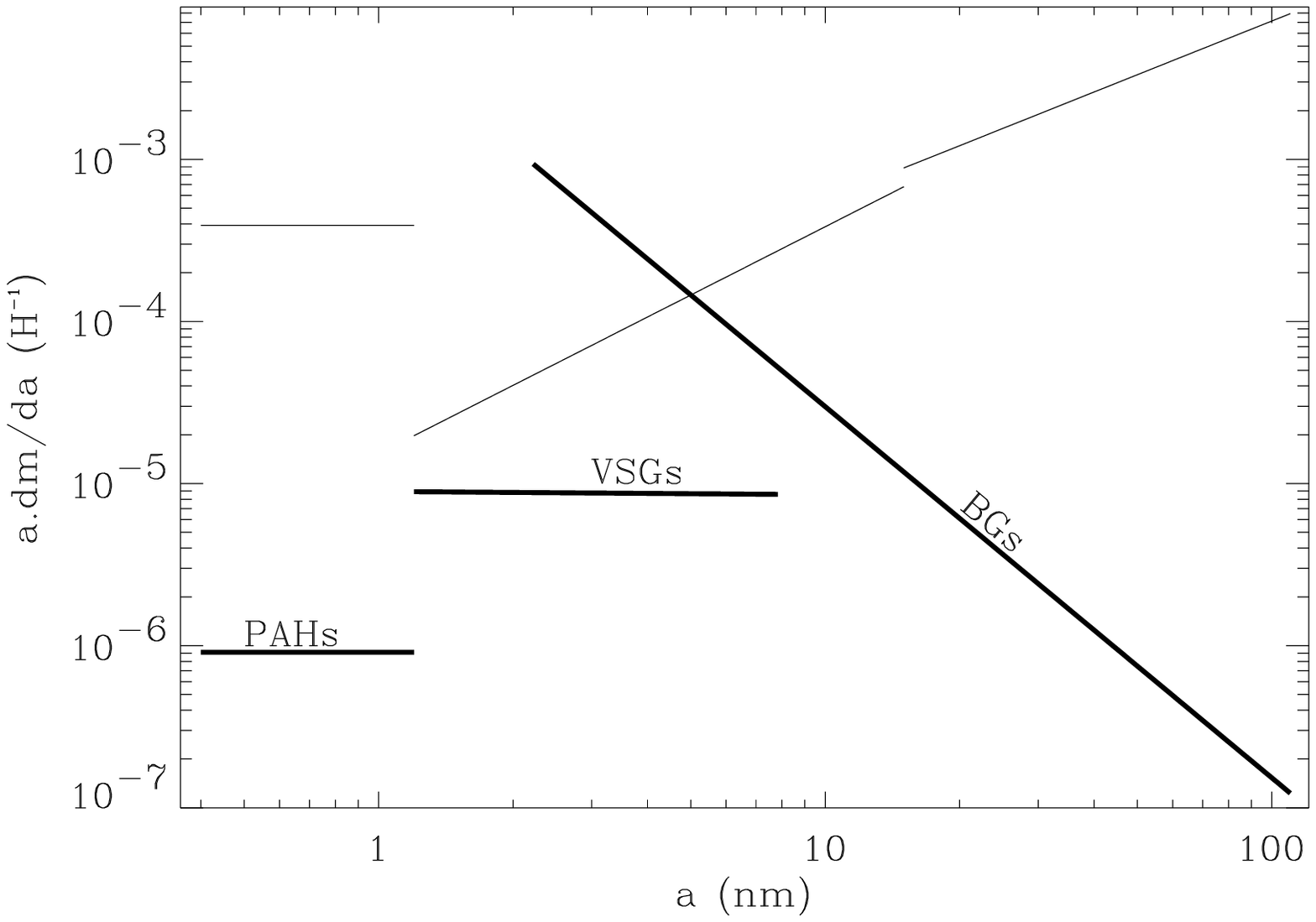}
  \caption{Size distribution of the grains in NGC~1569 and in the
           Galaxy.
           The three thin lines are the mass spectrum of the three dust
           components of the Galaxy.
           The bold lines are the mass spectrum of the three dust
           components of NGC~1569.
           $a$ is the grain radius and $m(a)$ is the dust mass of the component
           between $a$ and $a+da$.}
  \label{fig:sizedist}
  \end{center}
\end{figure}
Figure~\ref{fig:sizedist} compares the mass spectrum of the best solution of 
the 3 DBP90 dust components in NGC~1569 with that of the Galaxy. 
One significant difference is that the BGs in NGC~1569, are smaller overall 
compared to those of the Galaxy. 
The dust mass is concentrated in grains of smaller sizes in NGC~1569, 
$\sim 3\;\rm nm$, and the dust mass in large grains is almost negligible. 
In this case, what we called Very Small Grains and Big Grains have similar 
sizes but these names are maintained to be consistent with the original DBP90 
model.
However, we should call these two components carbonaceous grains and silicate 
grains respectively.

\begin{figure}[htb]
  \begin{center}
  \includegraphics[width=\hsize]{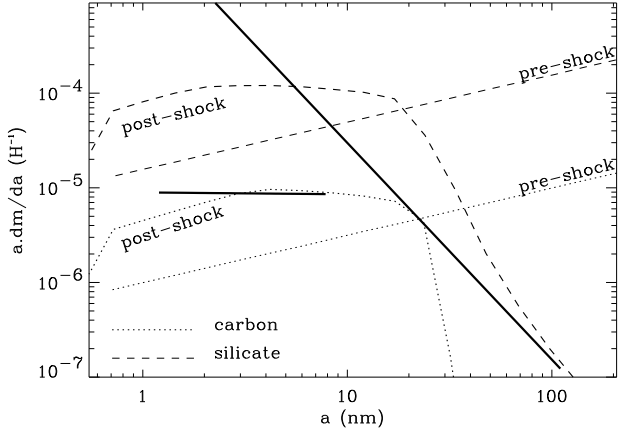}
  \caption{The effects of shocks on the dust size distribution.
           The solid lines are the mass spectrum of the VSGs which are
           carbonaceous grains and of the BGs which are silicates,
           in NGC~1569.
           The dotted lines show the mass spectra of carbon grains before
           (straight line) and after (curved line)
           a shock wave ($V_{\rm shock} = 100 \;\rm km\, s^{-1}$).
           The dashed lines represents the effect of the 
           $100 \;\rm km\, s^{-1}$ shock on silicate grains 
           (before shock: straight line and after shock: curved line).}
  \label{fig:shocks}
  \end{center}
\end{figure}

The size distribution of grains in NGC~1569 is consistent with that of grains 
which have been affected by shocks. 
Figure~\ref{fig:shocks} illustrates the effect of shocks on the size 
distribution of the carbon and silicate grains in NGC~1569 (Jones et al. 1996).
Both distributions have the same shape even though quantitatively, the cut-off 
does not occur at the same grain size. 
The disagreement of the shock model with the precise shape of the NGC~1569
dust size distribution reflects our ignorance of the initial size distribution
in NGC~1569, but the effect is qualitatively the same. 
The dust mass is transferred from large to small grains via shattering in 
grain-grain collisions in shock waves. 
NGC~1569 has recently experienced a large number of supernovae 
(Israel \& de Bruyn 1988; Waller 1991; Greve et al. 2002), thus 
supernova-generated shock waves should be common in the ISM of this galaxy.

The presence of smaller grain size distributions, that dominate the global IR
emission in dwarf galaxies, is opposite to the deduced dust properties in 
AGNs (e.g. Laor \& Draine 1993; Maiolino et al. 2001). 
In the case of AGNs, dust destruction effects due to sublimation or thermal 
sputtering can explain the absence of small grains in the hostile AGN 
environment, leaving only the larger, more robust grains. 
Although similar effects could be present in the hard radiation field of 
NGC~1569, the effects of shocks in redistributing the dust size distribution 
appears to be dominant. 
However, most of the dust mass is in the VCG component and we cannot yet 
deduce the size distribution of these grains, until such time as the 
submm/mm wavelength range of the SED is observed in more detail.
\sue{We give a rough estimate of the average size of the VCGs in 
Sect.~\ref{sec:explsub}.}

Due to the small grain sizes in the PAH, VSG and BG components, most of the
grains are stochastically heated and are not in thermal equilibrium with the
radiation field.  
%CHANGE 
\jpb{Table~\ref{tab:temp} contains the temperature range of each component 
for the maximum and minimum sizes compared to Galactic values.}
%CHANGE
\begin{table}
\begin{center}
  \begin{tabular}{*{6}{l}}
    \hline
    \hline
               &              & \multicolumn{2}{c}{\bf NGC~1569}            & 
      \multicolumn{2}{c}{\bf Milky~Way}           \\
               &              & $\pmb{\rm T_{min}}$  & $\pmb{\rm T_{max}}$  & 
      $\pmb{\rm T_{min}}$  & $\pmb{\rm T_{max}}$  \\
    \hline
      \bf PAH  & $\pmb{a_-}$  & 2.7~K                & 9200~K               & 
      2.7~K                & 4400~K               \\
               & $\pmb{a_+}$  & 2.7~K                & 1800~K               & 
      2.7~K                & 1100~K               \\
    \hline
      \bf VSG  & $\pmb{a_-}$  & 2.7~K                & 890~K                & 
      2.7~K                & 630~K                \\
               & $\pmb{a_+}$  & 2.7~K                & 160~K                & 
      2.7~K                & 78~K                 \\
    \hline
      \bf BG   & $\pmb{a_-}$  & 2.7~K                & 230~K                & 
      15~K                 & 22~K                 \\
               & $\pmb{a_+}$  & 28~K                 & 28~K                 & 
      17~K                 & 17~K                 \\
    \hline
  \end{tabular}
  \caption{Temperatures of the grains (other than VCGs).
           We give the minimum ($\rm T_{min}$) and maximum ($\rm T_{max}$) 
           temperatures for the minimum and maximum sizes, component by 
           component.
           All the grains are stochastically heated except the largest
           BGs where $\rm T_{min} = T_{max} = T_{equilibrium}$.}
  \label{tab:temp}
\end{center}
\end{table}
Stochastic heating is a very important process that must be taken into 
consideration in dust models applied to dwarf galaxies. 
Using models which assume modified black bodies for dust emission even up to 
FIR wavelengths in dwarf galaxies, when the grains can be thermally 
fluctuating, can result in an {\it underestimate} of the dust mass.
\ed{We have compared the grain radii at the transition between thermal
equilibrium and stochastically radiating dust ($a=a_t$) in NGC~1569 and in 
the Galaxy.
First, we compute the average energy of a photon, $\bar{\epsilon_\gamma}(a)$, 
\begin{equation}
  \displaystyle \bar{\epsilon_\gamma}(a) = 
     \frac{\displaystyle\int_0^\infty u_\nu Q_\nu(a)\: d\nu}{\displaystyle
           \int_0^\infty (u_\nu/h\nu) Q_\nu(a)\: d\nu}\; ,
\end{equation}
where $u_\nu=4\pi I_\nu/c$, $Q_\nu(a)$ is the absorption efficiency of a 
grain of radius $a$ and $c$ is the speed of light.
Second, we calculate the temperature of the dust, $T_d$, following a single
absorption of a photon of energy $\bar{\epsilon_\gamma}(a)$:
\begin{equation}
  \displaystyle \bar{\epsilon_\gamma}(a) = \frac{4\pi a^3}{3}
     \int_0^{T_d}C(T)\: dT \; ,
\end{equation}
using the expression for the specific heat capacity, $C(T)$, for silicate 
grains given by Draine \& Anderson (1985).
The transition occurs for radii where the cooling time of the dust, 
$\Gamma_{\rm cool}$, is equal to the photon absorption time, 
$\Gamma_{\rm abs}$, which can be estimated as follows:
\begin{equation}
  \left\{
  \begin{array}{l}
    \displaystyle 
      \Gamma_{\rm abs} \simeq 
      4\pi a^2\int_0^\infty 
      \frac{\pi I_\nu}{\bar{\epsilon_\gamma}(a)}Q_\nu(a)\: d\nu \\
    \\ \displaystyle 
      \Gamma_{\rm cool} \simeq
      4\pi a^2\int_0^\infty 
      \frac{\pi B_\nu(T_d)}{\bar{\epsilon_\gamma}(a)}Q_\nu(a)\: d\nu
  \end{array} 
  \right. \; .
  \label{eq:rates}
\end{equation}
$I_\nu$ is the intensity from Fig.~\ref{fig:compISRF}.
$\Gamma_{\rm cool}$ and $\Gamma_{\rm abs}$ are plotted in 
Fig.~\ref{fig:stochast} for NGC~1569 and the Galaxy.
\begin{figure}[h]
  \begin{center}
    \includegraphics[width=\hsize]{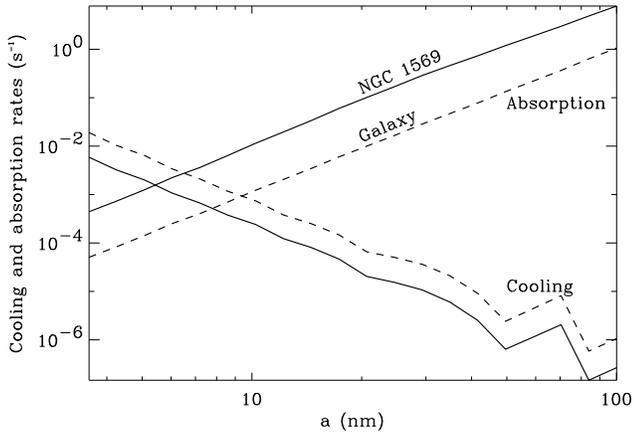}
    \caption{Cooling and photon absorption times computed using
             Eq.~\ref{eq:rates}.
             The solid lines are for NGC~1569 and the dashed lines for the 
             Galaxy.}
    \label{fig:stochast}
  \end{center}
\end{figure}
We find that the transition occurs at $a_t \simeq 5$~nm in NGC~1569 and 
$a_t \simeq 9$~nm in the Galaxy.
In other words, grains larger than $a_t$, for a given ISRF, will be
in thermal equilibrium with the radiation field and smaller than $a_t$, will 
be stochastically heated.
Alternatively, we can show that the order of magnitude of $a_t$ in the Galaxy 
is roughly twice $a_t$ in NGC~1569.
If we assume that $\bar{\epsilon_\gamma}$ is the same in both cases and 
that we are in the regime where $Q_\nu(a)/a$ is independent of $a$, then
we see that $\Gamma_{\rm abs} \propto a^3 L_{\rm ISRF}$ ($L_{\rm ISRF}$
being the integrated luminosity of the ISRF) and that $\Gamma_{\rm cool}$
depends only on $a$.
From the radiation fields in Fig.~\ref{fig:compISRF}, we know that 
$L_{\rm ISRF}$ is roughly 10 times larger in NGC~1569 than in the Galaxy.
Consequently, we expect that the transition radius in the Galaxy would be 
$10^{1/3}\simeq 3$ times the transition radius in NGC~1569, which is the 
correct order of magnitude.
This result means that grains are stochastically heated at larger sizes 
in the Galaxy than in NGC~1569.
However, the size distributions are different.
Thus, in NGC~1569, most of the grains are stochastically heated since the dust
mass (PAH, VSG and BG) is concentrated in small sizes and in the Galaxy, most 
of the emission originates in grains reaching thermal equilibrium since the 
dust mass is concentrated in large grains.}

\begin{figure*}[ht]
  \begin{center}
  \includegraphics[width=\textwidth]{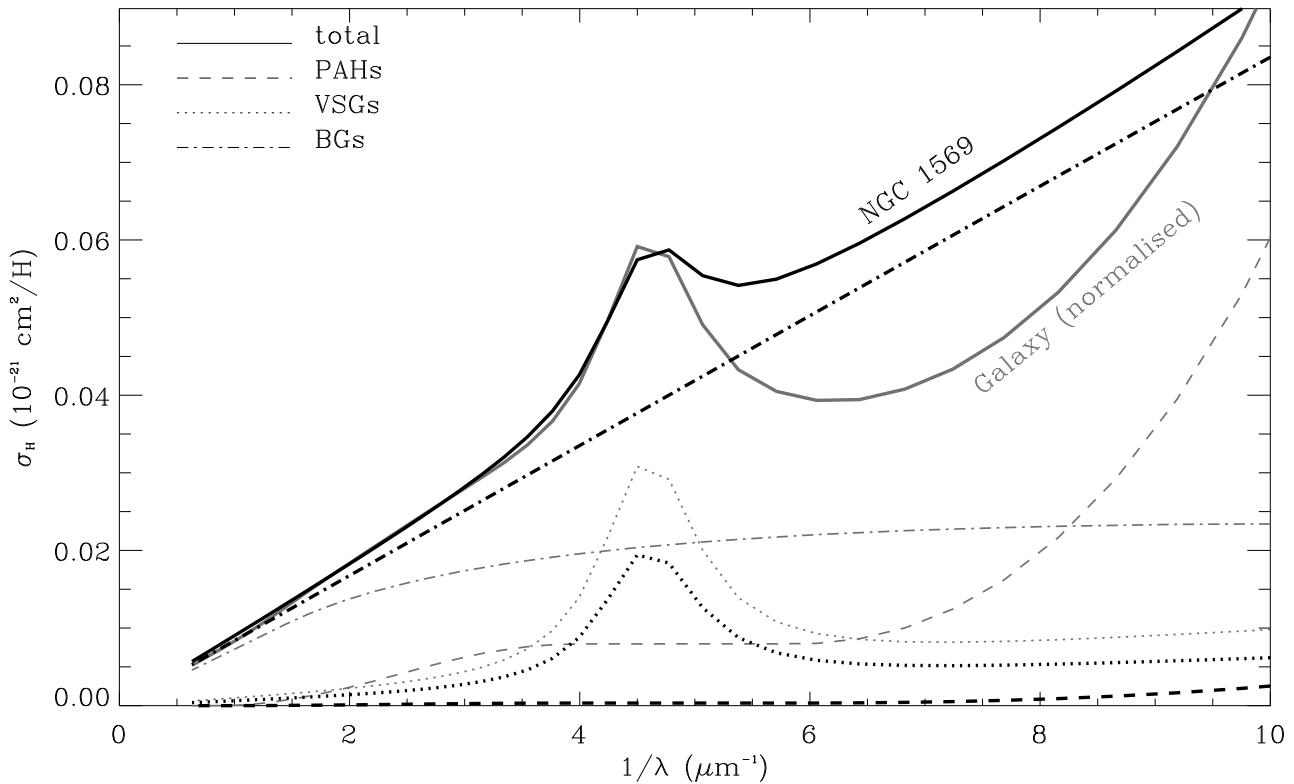}
  \caption{Extinction curve for NGC~1569 (black) compared to the Galaxy (grey,
           \chris{from DBP90}).
           The Galactic extinction curve has been scaled down for comparison 
           to that of NGC~1569.
           The solid lines are the total extinction curve modeled with DBP90.
           The contributions to the extinction curves from the individual dust
           components are also shown in dashed lines (PAHs), dotted lines 
           (VSGs) and dashed-dotted lines (BGs).
           The opacity is expressed as the cross section per H atom.}
  \label{fig:extinction}
  \end{center}
\end{figure*}

% Courbe d'extinction
%--------------------
As a result of the difference in the grain size distribution, the extinction 
curve of NGC~1569 is noticeably different from that of the Galaxy 
(Fig.~\ref{fig:extinction}).
This extinction curve consists only of the three standard DBP90 components and 
does not take into account the extinction due to the VCGs since we do not know
the size distribution of this dust component.
However, the energy emitted by the VCGs is very weak ($\sim 0.02\,\%$ of the
total energy radiated by the dust from MIR to mm wavelengths) thus the 
extinction due to this component should be very weak also.
The magnitude of the extinction is lower in NGC~1569 than the Galaxy at all 
wavelengths. 
Our iterative process gives $A_V = 0.45 \pm 0.05$. 
This $A_V$ corresponds to the case where all the dust mass of the three DBP90
model dust components is located in front of the stars.
The effective $A_V$ deduced from the energy balance, is
$A_V^{\rm eff} = 0.2$, less than the value $A_V = 0.65 \pm 0.04$, deduced by
Devost et al. (1997) from the $\rm H\alpha/H\beta$ line ratio toward the main
body of the galaxy.
However, they assumed Galactic extinction properties to derive this value.
Figure~\ref{fig:extinction} clearly shows that if we normalise the extinction
curves of the Galaxy and NGC~1569 at the V wavelength, the UV slopes are
very different indicating higher energy absorption in NGC~1569.
Thus, to obtain the same energy absorption, the $A_V$ adopted with the Galactic
properties should be higher, which is consistent with Devost et al. (1997).
\sue{In a general way, using Galactic extinction properties instead of the 
synthesized extinction curve of NGC~1569, when both are normalised by 
$\rm A_V$, gives an erroneously lower $\tau_{\rm UV}$ by a factor as low as 
$\sim 0.7$.}

\begin{figure}[h]
  \begin{center}
  \includegraphics[width=\hsize]{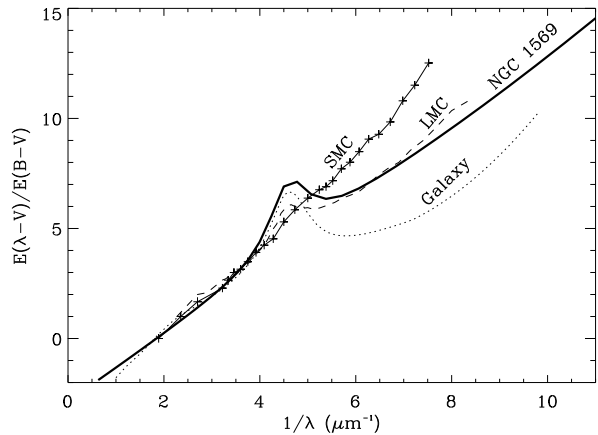}
  \caption{Comparison of various extinction curves.
           The Galactic curve is from Seaton (1979),
           the LMC's is the average of Koornneef \& Code (1981) and of
           Nandy et al. (1981) and
           the SMC's is from Pr\'evot et al. (1984).}
  \label{fig:compExtLMC}
  \end{center}
\end{figure}

We notice in Fig.~\ref{fig:extinction} that the lack of PAHs induces
a UV rise in the extinction curve that is quite linear, due to the
dominant effect of the BGs in the FUV rise and that the small size of
the more abundant grains (BGs) results in a relatively steeper
extinction curve. 
The bump at 2175~\AA\ is weaker in NGC~1569 than in the Galaxy, due to the 
dominance of the BGs over the VSG component. 
This peculiarity is also observed in the Large Magellanic Cloud (Fitzpatrick 
1985) extinction curve which has a metallicity similar to NGC~1569. 
Figure~\ref{fig:compExtLMC} compares the synthesized extinction curve for 
NGC~1569 to the observed extinction curves for the LMC (Koornneef \& Code 1981;
Nandy et al. 1981), the SMC (Pr\'evot et al. 1984) and the Milky Way (Seaton 
1979).
The similarity of the modeled extinction curve with that of the LMC is 
striking. 
However, the intensity of the bump is, of course, model dependent.  
Nevertheless, this bump should be weaker compared to the continuum since the 
slope of this continuum is steeper.

      % Masses de poussiere
      %--------------------
      \subsubsection*{The dust mass}

We determine the dust mass of NGC~1569 to be $(1.6 - 3.4)\times 10^5$~\msol, 
15 to 20 times more than that found previously by Lisenfeld \& Ferrara (1998),
who only had IRAS data available, and $\sim 4$ times more than that found by 
Lisenfeld et al. (2002) who use the DBP90 dust model, but assume Galactic dust 
grain properties and the Galactic ISRF.
Our results show that the BGs and VCGs dominate the dust mass, with 
very little contribution from the VSGs and PAHs,
40 to 70~$\%$ of the total dust mass resides in the VCG component. 
The total gas-to-dust mass ratio, $\mathcal{R}$, inferred for NGC~1569, 
assuming a total mass of hydrogen $\rm M_H = 1.9\times 10^8\; M_\odot$ 
(Israel 1997) and $\rm M_{He} = 0.25 M_{gas}$, is between $\mathcal{R}=740$ 
and $\mathcal{R}=1600$ (Table~\ref{tab:masses}). 
If we assume only that $\mathcal{R} \propto Z$, we should find, for NGC~1569,
$\mathcal{R}=770$ which is in agreement with our best $\chi^2$ value of
$\mathcal{R}=740$.
However if we use the law from Lisenfeld \& Ferrara (1998),
$1/\mathcal{R} \propto Z^{0.52 \pm 0.25}$, we find $\mathcal{R}=2900$ and a
lower limit of $\mathcal{R}=1200$ which would be correct if we did not have 
very cold dust.
It is not surprising since the model of Lisenfeld \& Ferrara (1998) does not 
take into account submillimeter data so they did not include very cold dust.
%CHANGE
\ed{We also deduce the dust-to-metal mass ratio 
$\mathcal{D} = 1/(1+\mathcal{R}Z)$ which is $\mathcal{D}\simeq 1/3$ in the 
Galaxy and $1/4\leq\mathcal{D}\leq 1/7$ in NGC~1569.
The smaller $\mathcal{D}$ value found in NGC~1569 could reflect the fact that
shocks can erode and destroy the dust, transferring more metals into the gas 
phase.}
%CHANGE
\begin{table*}[htbp]
\begin{center}
  \begin{tabular}{*{4}{l}}
    \hline
    \hline
      \multicolumn{1}{c}{}
      & \bf Dust mass         & \bf Gas-to-dust & \bf Dust-to-metal \\
      \multicolumn{1}{c}{}
      & \bf (\msol)           & \bf mass ratio  & \bf mass ratio    \\
    \hline
      \bf PAH
      & $\lesssim$ 190            &             &                   \\
      \bf VSG
      & $3.4\times 10^3$          &             &                   \\
      \bf BG
      & $8.4\times 10^4$          &             &                   \\
      \bf VCG
      & $(0.7 - 2.5)\times 10^5$  &             &                   \\
    \hline
      \bf TOTAL
      &                           &             &                   \\
      - without VCGs
      & $8.8\times 10^4$          & $2800$      & $\sim 1/11$       \\
      - with VCGs
      & $(1.6 - 3.4)\times 10^5$  & $740 - 1600$& $\sim 1/4 - 1/7$  \\
    \hline
  \end{tabular}
  \caption{Dust masses, gas-to-dust and dust-to-metal mass ratios in NGC~1569 
           deduced from the parameters in Table~\ref{tab:param}.
           \ed{The entries corresponding to non-relevant quantities are 
               blank for more clarity.}}
  \label{tab:masses}
\end{center}
\end{table*}

      % Discussion sur la distribution spatiale des poussieres
      %-------------------------------------------------------
      \subsubsection*{The spatial distribution of the dust
                      \chris{from the observations}}

Our multi-wavelength maps can be used to study the spatial
distribution of the various dust components.  
At first glance, the morphology of the submm/mm maps look similar to that of 
the \chris{ISOCAM-LW3 $15\; \rm \mu m$ and LW6 $7.7\; \rm \mu m$} (see
Fig.~\ref{fig:images}).
Lisenfeld et al. (2002) conclude that in NGC~1569 both the MIR and 
submillimetre/millimetre emission are due to the same dust component. 
Figure~\ref{fig:rapCAM_850} shows the spatial distribution of the ratio of 
\chris{LW6 (which traces the PAHs) to SCUBA-850~$\mu m$ (which traces the 
cold dust, both BG and VCG) and
the ratio of LW3 (which traces the VSGs) to SCUBA-850~$\mu m$ (cold dust)}. 
These ratio maps do not show a flat distribution. 
They demonstrate that the distribution of the hot dust emission from the VSGs 
is definitely more concentrated toward the region of the two most prominent 
H$\alpha$ peaks whereas the cold dust emission and the PAHs are more extended. 
The PAH emission is more extended than the VSG's but less so than the cold
dust emission. 
This is consistent with the simple view of an HII
region and a surrounding photodissociation region/molecular cloud. 
Hot dust emitting at $15\; \rm \mu m$ is seen to peak inside HII regions
in the Galaxy, while the PAHs peak at the photodissociated edges of
molecular clouds (e.g. Cesarsky et al. 1996; Abergel et al. 2000;
Klein et al. 1999). 
As the colder dust will be excited by similar sources, the morphology of the 
submm maps will appear to resemble the star formation tracers at these spatial
scales. 
In addition, the lower metallicity ISM of NGC~1569, tends to be clumpier, due 
to the lower dust abundance and, hence, larger mean-free path of photons and, 
subsequently, larger photodissociation effects.  
This clumpy effect means that the pervasive non-ionising radiation will be 
effective at exciting the surrounding colder dust. 
We find that, contrary to commonly-made assumptions, the bulk of cold dust in 
dwarf galaxies is not necessarily concentrated toward the outer regions, but 
is distributed between the star formation sites.

    %----------------------------------------------------------------------
    %                             L'ISRF
    %----------------------------------------------------------------------

  \subsection{The synthesized ISRF}
    \label{sec:synthISRF}

\begin{figure}[h]
  \begin{center}
  \includegraphics[width=\hsize]{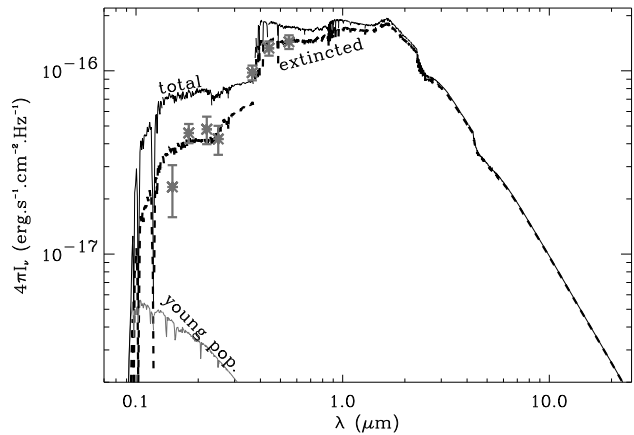}
  \caption{Synthesized ISRF for NGC~1569 computed with P\'EGASE and CLOUDY.
           The Points with error bars are the observational data from
           Table~\ref{tab:UVflux}, the solid black line is the global
           non-extincted ISRF, the dashed line is the global extincted ISRF
           and the grey line is the young single-burst component.
           The extinction curve used is the output from the dust model DBP90.}
  \label{fig:isrf}
  \end{center}
\end{figure}

The best synthesized global ISRF for NGC~1569 (Fig.~\ref{fig:isrf}) is a 
combination of two single-bursts with ages 4 and 100 Myr. 
The young component represents the recent starburst which has produced the 
two super-star-clusters (SSCs). 
These SSCs have populations of ages $4 - 7\; \rm Myr$ (Origlia et al. 2001), 
consistent with our younger population. 
The older component we use represents the underlying population present 
before the more recent starburst.
Although other older populations might be present, they do not dominate the 
global ISRF for this galaxy.

\begin{figure}[h]
  \begin{center}
  \includegraphics[width=\hsize]{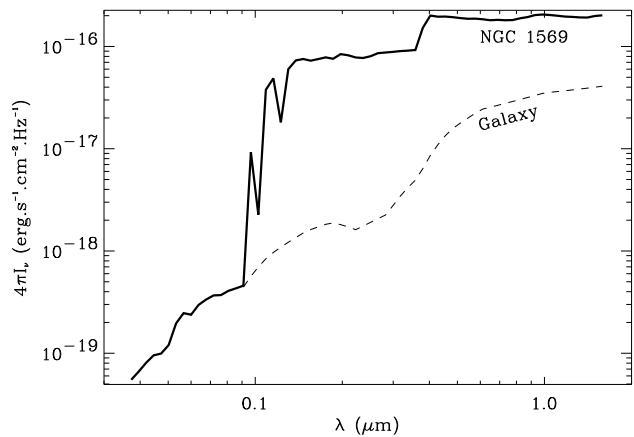}
  \caption{Synthesized ISRF for NGC~1569 (solid line) compared to the Galaxy 
           (dashed line).}
  \label{fig:compISRF}
  \end{center}
\end{figure}

% Comparaison avec la Voie-Lactee
%--------------------------------
Compared to the Galaxy, the global ISRF modeled for NGC~1569 is much harder in 
the UV range due to the contribution from the very young population 
(figure~\ref{fig:compISRF}). 
At $0.1\; \rm \mu m$, the ISRF from NGC~1569 is higher by more than an order 
of magnitude compared to the Galaxy. 
The ISRF of the Galaxy and other spiral galaxies is dominated by the cooler 
disk material while in NGC~1569, and other starbursting dwarfs, it is the star
formation activity that dominates the global ISRF.
This is also evident from the global MIR characteristics (Madden et al. 2003) 
and is also the case for other dwarf galaxies (Galliano et al. 2003). 
Therefore, when modeling the dust emission from dwarf galaxies, it is not 
valid to assume the Galactic ISRF. 

    %----------------------------------------------------------------------
    %                        AUTRES EXPLICATIONS
    %----------------------------------------------------------------------

  \subsection{An explanation for the millimetre excess}
  \label{sec:explsub}

% Intro
%------
There are large uncertainties in the physical properties of the VCGs. 
Such a large mass of dust locked up in very cold dust is not unlikely. 
However, there are some questions to address concerning the nature of these 
dust grains. 
How can a population of such very cold dust exist on galactic scales? 
How can this dust be so cold? 
Is very cold dust seen in other galaxies? 
In this section, we explore various hypotheses to explain the millimetre 
excess.

% Eventuel probleme observationnel
%---------------------------------
We can also invoke a non-dust effect (i.e. molecular line and/or radio 
continuum contamination) or a cross-calibration problem. 
The non-dust effects have already been taken into account (see 
Sect.~\ref{sec:scuba}).
Moreover the SED of II~Zw~40 (Galliano et al. 2003) shows the same trend in the
millimetre regime, perhaps even suggesting that this excess might be a general 
feature of starbursting dwarf galaxies.

    % Exemples de poussiere froide dans la litterature
    %-------------------------------------------------
    %CHANGE********
    %\subsubsection{Cold dust in the literature}
    \subsubsection{Very cold dust in the literature}

% Poussiere froide
%-----------------
A very cold component has been suggested for the Galaxy. 
For example, Reach et al. (1995) found a component with $\rm T=5-7$~K in some 
Galactic continuum spectra. 
%CHANGE********
\jpb{ 
More recently, Boulanger et al. (2002) published COBE and Archeops data for 
the Galaxy that implies a $\sim 5$~K dust component (with $\beta = 2$).
This result has not yet been independently confirmed.
This component of very cold dust appears to follow the HI gas, suggesting 
that it is associated with the cold neutral medium.
%More recently, Boulanger et al. (2002) find a very cold dust component in the 
%Galaxy from COBE and Archeops data that implies a $\sim 5$~K dust component
%(with $\beta = 2$).
%This component of very cold dust, however, appears to follow the HI gas, 
%suggesting that it is associated with the cold neutral medium.
Dupac et al. (2002) reported PRONAOS observations of the M~17 complex and 
found dust as cool as 10~K in some cold clumps.
}
Popescu et al. (2002) observed a sample of 63 late-type galaxies in
the Virgo cluster using ISOPHOT. 
Although their wavelength coverage is limited, observing only up to 
$170\; \rm \mu m$, they find a cold dust component is present in most cases,
with temperatures as low as 10~K in irregular and blue compact 
dwarf galaxies.

% Estimation de James et al. 2002
%--------------------------------
James et al. (2002) calculated dust masses for galaxies from the submillimeter 
emission, assuming that the fraction of metals incorporated in the dust in 
galaxies is a universal constant. 
For NGC~1569 they determine a dust mass of $\sim 1 \times 10^5$~\msol, 
which is similar to the mass that we derive ($1.6 - 3.4\times 10^5$~\msol). 
% .. repenser au modele de Eli (mais il faut connaitre l'histoire de SF de la 
% galaxie).

% Comparaison avec Ute
%---------------------
Lisenfeld et al. (2002) published a dust SED for NGC~1569 computed with
the standard DBP90 model.
They found no submillimetre excess and discount the presence of very cold 
dust since they were able to explain the submm emission with the VSGs.
We now discuss the reasons for our different results.
We find that the use of the CVF spectrum in the MIR provides an important and 
tight constraint on the slope of the VSG size distribution in our model.
Lisenfeld et al. (2002), however, characterise the MIR wavelength regime with 
only a single 12~$\mu m$ IRAS data point.
%Their SCUBA fluxes are obtained from a different observing run and the flux of
%their 450~$\mu m$ data is higher and has a larger error bar than those 
%reported here (Sect.~\ref{sec:obs}).
This effects the fit significantly.

In addition to the different observational constraints used to construct the 
observed SED, Lisenfeld et al. (2002) used the DBP90 model in a very 
different, more limited way than that presented here.
Since we have more data to constrain the DBP90 model, we allow the grain 
properties to differ from those of the original DBP90 model, which was 
constructed to explain the Galactic dust emission.
Lisenfeld et al. (2002) use the original Galactic dust parameters to obtain 
their fit.
Their input ISRF is that of the Galaxy, which is much softer than that of 
NGC~1569 (see section~\ref{sec:synthISRF}), and is scaled up by a factor of 
$\sim 60$.
They did not investigate the dust size distribution parameter space.
Our results are based on a $\chi^2$ evaluation to obtain the best fit.
Lisenfeld et al. (2002) conclude that the VSG component alone accounts for the 
FIR and submm and mm continuum.
In our model, such a solution was attempted, but resulted in a very poor
$\chi^2$ since the MIR observations could not be well-fitted.
% Il n'ont pas applique la conservation de l'energie... pitetrekeujleudihoupa?

    % Changement des proprietes optiques
    %-----------------------------------
    \subsubsection{Optical properties of BGs}

% Proprietes optiques des BGs
%----------------------------
We have investigated other means to increase the emissivity in the submm, 
instead of invoking a VCG component.
A change in the emissivity slope could be due to a change in the optical
properties of the big grains.  
Indeed, Mennella et al. (1998) measure the absorption coefficient per unit 
mass ($Q_{abs}$) of cosmic dust analog grains, over the temperature range 
$24-295$~K.  
The laboratory experiments, show a decrease of the emissivity index, 
$\beta$, as the temperatures increases, effecting the $Q_{abs}$ for wavelengths
$\lambda > 30\; \rm \mu m$.  
Our big grains reach higher temperatures than those of the Galaxy 
(Table~\ref{tab:temp}) and, in the standard model, 
$Q_{abs}^{\rm BG} \propto \lambda^{-2}$ for $\lambda > 100\;\rm \mu m$.
We attempted  to roughly reproduce the effect observed by Mennella et al. 
(1998) using $Q_{abs}^{\rm BG} \propto \lambda^{-\gamma}$ for 
$\lambda > 100\;\rm \mu m$ with $\gamma < 2$.
However, this modification failed to give an acceptable fit to the data in the 
submm/mm wavelength range.

Grain-grain coagulation, leading to porous aggregates, can increase dust 
emissivities at submm wavelengths. 
This has been invoked to explain the elevated opacity of circumstellar dusty 
envelopes and dense molecular cores (Ossenkopf \& Henning 1994; 
Stepnik et al. 2003). 
However, this effect is localised and has not yet been seen on galactic scales,
as would be necessary to explain the data presented here.

    % Transfert radiatif
    %-------------------
    \subsubsection{Embedded dust as the source of the mm excess}

% Utilisation generale du modele
%-------------------------------
The properties of the dust differ considerably from one region to another in 
a galaxy, depending on the nature of the environment (HII regions, cirrus, 
dense core, etc.).  
The DBP90 model has been constructed to describe the diffuse ISM of the Galaxy.
Consequently, applying it to an entire galaxy like NGC~1569 is not
straightforward.  
However, NGC~1569 is a peculiar galaxy: it is a starbursting dwarf and its 
emission properties show that it resembles a giant HII region (particularly in 
the MIR; see Madden et al. 2003).
% Non-uniformite de la distribution
%----------------------------------
Moreover, the emission from hot dust is enhanced relative to the cold dust 
in the starburst regions (see Fig.~\ref{fig:rapCAM_850}).
This departure could explain our inability to fit the SED with only the
three standard components.
The various phases of the ISM could have very different temperatures and 
technically should not be treated as if there were the same physical region.

\begin{figure}[h]
  \begin{center}
  \includegraphics[width=\hsize]{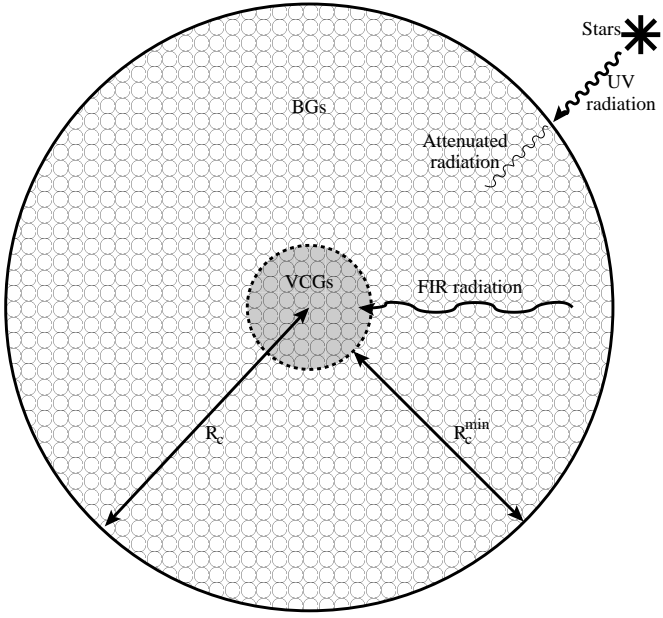}
  \caption{Schematic diagram of a clump.}
  \label{fig:clump}
  \end{center}
\end{figure}

% On s'amuse avec les ordres de grandeur
%---------------------------------------
If we assume that the very cold dust is made up of grains deeply embedded in
clouds, where the UV-optical radiation can not easily penetrate, the dust 
FIR-mm emission could be a heating source for this dust component 
(Fig.~\ref{fig:clump}).
%CHANGE
We can make a rough estimate of the temperature of the very cold grains 
(i.e, the VCGs) to verify this explanation.
Let us assume that the photons which heat the VCGs are due to the emission 
from the BGs, 
simply modeled as a blackbody, made of grains of radius $\rm a_{BG}$, 
\ed{temperature $\rm T_{BG} = 57\;K$ (corresponding to the peak wavelength of 
the BG component), mass $\rm M_{BG}$ and emission 
coefficient $\rm Q_{em}(a_{BG},\nu_{BG})$, $\rm \nu_{BG}$ being the 
frequency where the maximum emission by the BGs occurs}.
\sue{These are the parameters that we derived from our fitting of the SED of
NGC~1569.}
The VCGs are modeled with a blackbody of
\ed{temperature $\rm T_{VCG}$, mass $\rm M_{VCG}$ and
absorption and emission coefficients $\rm Q_{abs}(a_{VCG},\nu_{BG})$, 
$\rm Q_{em}(a_{VCG},\nu_{VCG})$, 
$\rm \nu_{VCG}$ being the frequency where the maximum emission by the 
VCGs occurs and are made of grains of radius $\rm a_{VCG}$ embedded in a clump
of radius $\rm R_{c}$.
The luminosities emitted by these two components, $\rm L_{BG}$ and 
$\rm L_{VCG}$, can be determined from the number of grains of each 
species, $\rm N_x = 3 M_x / 4\pi a_x^3 \rho_x$ ($\rm x$ being either BG or VCG
and $\rm\rho_x$ being the specific mass densities),
and the luminosity radiated by only one grain 
$\rm \mathcal{L}_x = 4\pi a_x^2 Q_{em}(a_x,\nu_x) \sigma T_x^4$. 
The total luminosity is then $\rm L_x = N_x \mathcal{L}_x$ 
($\rm x$=BG or VCG).
The ratio $\rm \Theta = L_{BG} / L_{VCG}$ is
\begin{equation}
  \begin{array}{rl}
    \displaystyle \rm  
    \Theta \simeq & \displaystyle \rm  
               \left(\frac{M_{BG}}{M_{VCG}}\right) 
               \left(\frac{\rho_{VCG}}{\rho_{BG}}\right) 
               \left(\frac{a_{VCG}}{a_{BG}} \right) \times \\
  &     \displaystyle \rm  
     \left(\frac{Q_{em}(a_{BG},\nu_{BG})}{Q_{em}(a_{VCG},\nu_{VCG})}\right)
               \left(\frac{T_{BG}}{T_{VCG}}\right)^4
  \end{array}
  \label{eq:Theta}
\end{equation}
and its numerical value from the model (Fig.~\ref{fig:dustembb}) is 
$\Theta = 4.1 \times 10^4$.
Using the same method, the heating and cooling luminosities for the VCGs are:
\begin{equation}
  \left\{
  \begin{array}{l}
    \displaystyle  \rm
     L_{heat} \simeq \frac{3M_{BG}}{4\pi a^3_{BG}\rho_{BG}}\pi a_{BG}^2
                     Q_{em}(a_{BG},\nu_{BG})\sigma T_{BG}^4          \\
    \\
    \displaystyle  \rm
     L_{cool} \simeq \frac{3M_{VCG}}{4\pi a^3_{VCG}\rho_{VCG}} 4\pi a_{VCG}^2
                     Q_{em}(a_{VCG},\nu_{VCG})\sigma T_{VCG}^4
  \end{array}
  \right.
\end{equation}
which is equivalent to $\rm L_{heat} = 1/4 L_{BG}$ and 
$\rm L_{cool} = L_{VCG}$.
This factor of $1/4$ is due to the fact that BGs do not radiate only in 
the direction of the VCGs inside the clump, they also radiate away without 
intercepting the VCGs.
The powers absorbed and emitted by one VCG grain are:
\begin{equation}
  \left\{
  \begin{array}{l}
    \displaystyle  \rm
    P_{abs} \simeq \frac{L_{heat}}{4\pi R_{c}^2} 4\pi a_{VCG}^2
                   Q_{abs}(a_{VCG},\nu_{BG}) \\
    \\
    \displaystyle  \rm
    P_{em} \simeq \frac{L_{cool}}{4\pi R_{c}^2} 4\pi a_{VCG}^2
  \end{array}
  \right.\; .
\end{equation}
Equating absorption and emission gives:
\begin{equation}
  \begin{array}{rl}
    \displaystyle \rm
      T_{VCG} \simeq
    & \displaystyle \rm
      \left[
      \frac{1}{4}\,
      \left(\frac{M_{BG}}{M_{VCG}}\right)
      \left(\frac{a_{VCG}}{a_{BG}}\right)
      \left(\frac{\rho_{VCG}}{\rho_{BG}}\right) \times
    \right. \\
    & \displaystyle \rm \left.
      \left(\frac{Q_{em}(a_{BG},\nu_{BG})Q_{abs}(a_{VCG},\nu_{BG})}
                 {Q_{em}(a_{VCG},\nu_{VCG})}\right)
      \right]^{1/4} T_{BG} \; .
  \end{array}
\label{eq:Tvcg}
\end{equation}
In the Eq.~\ref{eq:Tvcg}, the two unknowns are $\rm T_{VCG}$ and
$\rm a_{VCG}$.
We can deduce $\rm a_{VCG}$ by coupling Eqs.~\ref{eq:Theta} and \ref{eq:Tvcg}:
\begin{equation}
  \rm \Theta \simeq 4 / Q_{abs}(a_{VCG},\nu_{BG}) \;.
  \label{eq:radius}
\end{equation}
To be consistent with our model, which assumes $\beta = 1$, the VCGs are 
carbonaceous, $\rm \rho_{VCG} = 2.2 \; g\, cm^{-3}$ and 
$\rm \rho_{BG} = 3.3 \; g\, cm^{-3}$.
To determine the order of magnitudes of the efficiencies, we refer to 
Draine \& Lee (1984).
For the BGs we use the astronomical silicate values
at $\lambda = 90\;\mu m$, $\rm Q_{abs}/a \simeq 0.017 \; \mu m^{-1}$.
This is in the regime where $\rm Q_{abs}/a$ does not depend on the grain 
radius.
For the VCGs, we used graphite values.
The coefficients are: 
$\rm Q_{abs}/a \simeq 0.022 \; \mu m^{-1}$ at $\lambda = 90\;\mu m$ and
$\rm Q_{abs}/a \simeq 4.5\times 10^{-4} \; \mu m^{-1}$ at 
$\lambda = 660\;\mu m$.  
Eq.~\ref{eq:radius} gives us the radius of the VCGs, $\rm a_{VCG} \simeq 4$~nm
is also in the same regime where $\rm Q_{abs}/a$ is independent of a.
With this simple analytical model, the temperature of the VCGs is estimated 
to be $\rm 6\;K \lesssim T_{VCG} \lesssim 9\;K$.}
Thus, the hypothesis of very cold dust hidden in clumps bathed by the FIR 
radiation from warmer grains is consistent and gives the appropriate order of 
magnitude of $\rm T_{VCG}$.
%CHANGE
\jpb{Moreover, Bernard et al. (1992), using a radiative transfer model 
coupled with DBP90 in the Galactic case, showed that the temperature of the 
dust deeply embedded in a dense cloud is $\sim 6$~K.}
Radiative transfer may indeed turn out to be the key to understanding this 
property of the dust spatial distribution in these galaxies.

%CHANGE
\ed{We can attempt to estimate the clump radius ($\rm R_c$).
First, we can determine a minimum clump radius ($\rm R_c^{min}$) by estimating
the optical depth required to shield the stellar radiation, since in the 
above scenario, the VCGs are heated only by the envelope of BGs.
We consider that the VCGs are screened by the BGs when the stellar energy
seen by the VCGs is at most equal to the energy emitted by the VCGs.
Given our SED and the extinction curve of Fig.~\ref{fig:extinction}, the 
optical depth in the V band needed is $\tau_V^{shield} \simeq 60$.
The average wavelength of the stellar radiation is 
$\rm \lambda_{ISRF} \simeq 0.2\; \mu m$.
At this wavelength, the optical depth needed is 
$\tau^{shield}(\rm\nu_{ISRF}) \simeq 200$.
At the wavelength where the BGs emit ($\rm\lambda = 90\; \mu m$), the optical
depth is $\tau=0.1$, thus we verify that the shell is optically thin to this 
radiation.
The contrast density between the clumps (index $c$ hereafter) and the 
interclump media (index $icm$ hereafter) is $\rm \alpha = n_c/n_{icm}$.
The minimum radius of a clump $\rm R_{c}^{min}$ is the radius required to 
reach the optical depth of $\tau^{shield}(\rm\nu_{ISRF})$.
If we assume that only the BGs are responsible for the shielding,
\begin{equation}
  \tau^{shield}(\rm\nu_{ISRF}) = \rm R_{c}^{min} n_{c}^{BG} \pi a^2_{BG}
       Q_{abs}(a_{BG},\nu_{ISRF})
  \label{eq:tauClump}
\end{equation}
where $\rm n_{c}^{BG}$ is the density of the BGs in the clumps and 
$\rm Q_{abs}(a_{BG},\nu_{ISRF}) \simeq 0.1$ from Draine \& Lee (1984).
The actual clump size is difficult to estimate without some assumptions for 
$\rm n_{c}^{BG}$ or $\alpha$.
Let us assume that $\rm n_{c}^{BG}$ is the homogeneous density 
$\rm n_{hom}^{BG} \simeq 3\times 10^{-8}$~cm$^{-3}$ scaled by $\alpha$.
From Eq.~\ref{eq:tauClump}, we find that $\rm R_{c}^{min} \simeq 8$~pc if
$\alpha = 10^4$, the extreme case, where there is only one clump, and 
$\rm R_{c}^{min} \simeq 0.8$~pc if $\alpha = 10^5$.
To estimate the size of the VCG core, we require that it is optically
thin to the BG radiation.
The ratio between the mass of BGs and VCGs constrains the density of this core.
The maximum radius of the VCG core is $\sim 4 - 6$~pc when $\alpha = 10^4$, 
the number of clumps is 1 or 2 and the volume filling factor 
$\phi \simeq 3 - 6 \times 10^{-5}$.
When $\alpha = 10^5$, the radius of the
core is $\sim 0.4 - 0.6$~pc, the number of clumps is between 200 and 300 and 
the filling factor $\phi \simeq 3 - 6 \times 10^{-6}$.
This rough estimation tends to imply that there would be a small number of 
dense clumps in this galaxy.
Berkuijsen (1999) gives a range of $10^{-3} - 10^{-5}$ for the H$_2$ volume
filling factor of the Galaxy which is greater or equal to the range we find in
NGC~1569.
This clumpy characteristic is also consistent with relatively bright 
158~$\mu m$ [CII] emission seen in NGC~1569 and in low-metallicity 
environments, in general (Jones et al. 1997).}

  %==========================================================================
  %                                 Resume
  %==========================================================================

\section{Summary and conclusions}
\label{sec:concl}

Our modeling of the dust SED in NGC~1569 suggests that the nature of the dust 
in low-metallicity environments differs from that of metal-rich galaxies. 
\begin{enumerate}
  \item We have presented new SCUBA images at 450 and 850~$\mu m$.
        With additional data from the literature we constructed the
        observed dust SED for this galaxy.
  \item We have used a stellar evolution model (PEGASE), a photoionisation 
        model (CLOUDY) and a dust model (DBP90) to compute a self-consistent
        theoretical dust SED, independently synthesizing a global ISRF from 
        UV-to-optical observations with further constraints from MIR-ionic 
        lines.
        The DBP90 model has been used to investigate a large range of dust
 	parameters, making the link with the extinction deduced from the dust
 	model and the deredenning of the UV-to-optical data.
  \item The synthesized global ISRF is consistent with a very young population
        produced by the recent starburst and an older population tracing the
        underlying stars of the galaxy.
  \item We find very low abundances of PAHs, and smaller overall sizes of 
        grains emitting in the MIR and FIR, compared to the Galaxy. 
        Due to the small sizes, most of the grains are in a stochastic heating
        mode, and not in thermal equilibrium with the radiation field - even 
        grains emitting at FIR wavelengths.
  \item The presence of a millimetre emission excess can be explained by 
        ubiquitous clumps of very cold (5 to 7 K) dust grains with $\beta = 1$.
        These very cold grains can be responsible for 40 to 70~$\%$ of the
        total dust mass in the galaxy, 
        $\rm M_{dust} = (1.6 - 3.4)\times 10^5$~\msol.
        \sue{The absence of important submm/mm observations can result in, at 
        least, an order of magnitude of dust mass being missed.}
  \item The gas-to-dust mass ratio ranges from 740 to 1600, greater than that 
        in the Galaxy, even taking into account the relatively large mass of
        very cold grains.
        The dust-to-metals mass ratio ranges from $1/4$ to $1/7$, smaller than
        the Galactic value.
  \item From the dust size distribution, we generate an extinction curve for
        NGC~1569 (which has similar characteristics to that of the LMC). 
        We also derive a synthesized radiation field and a 
        wide range SED (UV - mm), for the global galaxy.
  \item \sue{The bulk of the cold dust traced by the submm/mm observations is 
        distributed between star forming regions, not concentrated toward the 
        outer regions of the galaxy.
        Our results are consistent with a clumpy medium and a filling 
        factor lower than $10^{-4}$.}
\end{enumerate}

Due to the low metallicity of NGC~1569, the ISM would appear to be very 
clumpy, and this has important consequences for the dust models. 
From our detailed dust modeling of dwarf galaxies (see also Galliano et al.
2003), we caution that dust models assuming the Galactic dust size 
distribution, Galactic extinction curve and Galactic ISRF will be in error.  
Models not dealing with stochastic heating processes should be dealt with 
cautiously, as this is probably the most important physical process for dust 
heating in dwarf galaxies.  
Another caution that follows from this work is the importance of obtaining 
data longward of the FIR. 
This is presently easier said than done.
However, in the near future we will have the capability to sample the submm/mm
emission from galaxies in more detail, and with greater sensitivity, with 
SOFIA, ASTRO-F, Herschel, Planck and ALMA. 
Without a proper understanding of the submm/mm SED of galaxies, galaxy number 
counts could be incorrect since the excess submm/mm emission might erroneously 
be attributed to higher star formation activity in redshifted galaxies, 
instead of the presence of a large, very cold dust mass.

  %==========================================================================
  %                          Remerciements
  %==========================================================================

\begin{acknowledgements}
We would like to thank Ren\'e Gastaud, H\'el\`ene Roussel, Pierre Chanial and
Marc Sauvage for their expert advice on ISOCAM data reduction;
Jean-Luc Starck for useful discussions on data processing techniques;
Bertrand Stepnik for his help on ISOPHOT data reduction and
Martin Haas and Ulrich Klaas for their advise concerning the use of
ISOPHOT data. 
We also thank David Hollenbach, Fran\c{c}ois Boulanger, Ute Lisenfeld and 
Frank Israel for helpful scientific discussion.
We are grateful to the referee, Eli Dwek, for his erudite comments that
helped to improve the quality of the paper.
\end{acknowledgements}

  %==========================================================================
  %                          Bibliographie
  %==========================================================================

  %==========================================================================

\end{document}